\documentclass[pra,a4paper]{revtex4}
\usepackage{amsthm}
\usepackage{bbm}
\usepackage{amssymb}
\usepackage{amsmath}
\usepackage{verbatim}
\usepackage{graphicx}
\usepackage{epsfig}

\usepackage{pb-diagram}
\usepackage{lamsarrow}
\usepackage{pb-lams}

\newtheorem{thm}{Theorem}
\newtheorem*{thm*}{Theorem}
\newtheorem{prop}[thm]{Proposition}
\newtheorem{cor}[thm]{Corollary}
\newtheorem{lem}[thm]{Lemma}
\theoremstyle{definition}

\newtheorem{obs}[thm]{Remark}

\theoremstyle{remark}

\newcommand{\lra}{\longrightarrow}

\newcommand{\w}{\omega}
\newcommand{\ten}{\otimes}

\newcommand{\R}{\mathbb{R}}
\newcommand{\C}{\mathbb{C}}
\renewcommand{\P}{\mathbb{P}}

\newcommand{\U}{\mathcal{U}}
\newcommand{\I}{\mathcal{I}}
\newcommand{\B}{\mathcal{B}}
\newcommand{\1}{\mathbbm{1}}

\newcommand{\la}{\lambda}

\newcommand{\ez}{\mathbb E}
\renewcommand{\>}{\rangle}
\newcommand{\<}{\langle}

\DeclareMathOperator{\id}{id} \DeclareMathOperator{\tr}{tr}

\DeclareMathOperator{\spanned}{span}

\DeclareMathOperator{\re}{Re} \DeclareMathOperator{\im}{Im}
\DeclareMathOperator{\gl}{gl} \DeclareMathOperator{\rc}{RC}

\begin{document}

\title{Unbounded violation of tripartite Bell inequalities }

\author{D. P\'{e}rez-Garc\'{\i}a}
\affiliation{Departamento de An\'{a}lisis Matem\'{a}tico,
Universidad Complutense de Madrid, 28040, Madrid, Spain }
\author{M.M. Wolf}
\affiliation{Max Planck Institut f\"{u}r Quantenoptik,
Hans-Kopfermann-Str. 1, Garching, D-85748, Germany}
\author{C. Palazuelos}
\affiliation{Departamento de An\'{a}lisis Matem\'{a}tico,
Universidad Complutense de Madrid, 28040, Madrid, Spain}
\author{I. Villanueva}
\affiliation{Departamento de An\'{a}lisis Matem\'{a}tico,
Universidad Complutense de Madrid, 28040, Madrid, Spain}

\author{Marius Junge}
\affiliation{Department of Mathematics, University of Illinois at
Urbana-Champaign, Illinois 61801-2975, USA}

\begin{abstract}
We prove that there are tripartite quantum states (constructed from
random unitaries) that can lead to arbitrarily large violations of
Bell inequalities for dichotomic observables. As a consequence these
states  can withstand an arbitrary amount of white noise before they
admit a description within a local hidden variable model. This is in
sharp contrast with the bipartite case, where all violations are
bounded by Grothendieck's constant. We will discuss the possibility
of determining the Hilbert space dimension from the obtained
violation and comment on implications for communication complexity
theory. Moreover, we
show that the violation obtained from generalized GHZ states is
always
bounded so that, in contrast to many other contexts, GHZ
states do in this case not lead to extremal quantum correlations.
In order to derive all these physical consequences, we will have to obtain new mathematical results in the
theories of operator spaces
and tensor norms. In particular, we will prove the existence of
bounded but not completely bounded trilinear forms from commutative
C*-algebras.
\end{abstract}

\date{\today}

\maketitle

\section{Introduction}

Bell inequalities characterize the boundary of correlations
achievable within classical probability theory under the assumption
that Nature is local \cite{WernerWolf}. Originally, Bell \cite{Bell}
proposed the inequalities, which now bear his name, in order to put
the intuition of Einstein, Podolski and Rosen \cite{EPR} on
logically firm grounds, thus proving that an apparently metaphysical
dispute could be resolved experimentally. Nowadays, the verification
of the violation of Bell inequalities has become experimental
routine \cite{Aspect,Rowe,Aspel}(albeit there is a remaining desire
for a unified loophole-free test). On the theoretical side---in the
realm of quantum information theory---they became indispensable
tools for understanding entanglement
\cite{Werner,entmulti1,entmulti2,entmulti3} and its applications in
cryptography \cite{Ekert,Acin0,Acin01,Acin1,Acin02} and
communication complexity \cite{CCP}. In fact, the insight gained
from the violation of Bell inequalities enables us even to consider
theories beyond quantum mechanics \cite{Masanes1,Wernercloning} and
allows to replace quantum mechanics by the violation of some Bell
inequality in the set of trusted assumptions for secure
cryptographic protocols
\cite{Acin0,Acin01,Acin02,Kent,MasanesWinter}.

Most of our present knowledge on Bell inequalities and their
violation within quantum mechanics is based on the paradigmatic
Clauser-Horne-Shimony-Holt (CHSH) inequality \cite{CHSH}. It bounds
the correlations obtained in a setup where two observers can measure
two dichotomic observables each. In fact, it is the only non-trivial
constraint on the polytope of classically reachable correlations in
this case \cite{Fine}. If we allow for more observables (measurement
settings) per site or more sites (parties) the picture is much less
complete. Whereas for two dichotomic observables per site the
complete set of multipartite `full-correlation inequalities' and
their maximal violations within quantum mechanics is still known
\cite{WW,ZB}, the case of more than two settings is, despite
considerable effort
\cite{moresettings1,moresettings3,moresettings2}, largely
unexplored.

One reason is, naturally, that
finding all possible Bell inequalities is a computationally hard
task
\cite{Pit1, Naor} and that in addition the violating quantum systems
become vastly more complicated as the number of sites and dimensions
increases. Another reason could be the lack of appropriate
mathematics to tackle the problem. Thus far, researchers have
primarily used algebraic and combinatorial techniques.

In this work, following the lines already implicit in
\cite{Tsirelson}, we will relate tripartite Bell inequalities with
two powerful theories of mathematical analysis: operator spaces, and
tensor norms. We will give new mathematical results inside these
theories and show how to apply them to provide a deeper insight into
the understanding of Bell inequalities, by proving some new and
intriguing
results on their maximal quantum violation. It
is interesting to note here that operator spaces have recently also
led to other applications in Quantum Information
\cite{Marius}.

We will start by outlining the main result and some of its
implications within quantum information theory. Sec. \ref{sec:preliminaries} will then
recall basic notions from the theory of operator spaces and
tensor norms and bridge between the language of Bell inequalities
and the mathematical theories. In Sec. \ref{sec:results2} we will prove that the violation remains
bounded for GHZ states. Finally, Sec. \ref{sec:proofs} provides the proof for the main
theorem.

\section{Main Result and Implications}\label{sec:Mainresult}

We begin by specifying the framework. For the convenience of the
non-specialist reader we will give first a brief introduction to
Bell Inequalities. For further information we refer the reader to
\cite{WernerWolf}.

\

Bell inequalities can be dated back to the famous critic of
Quantum Mechanics due to Einstein, Podolski and Rosen \cite{EPR}.
This critic was made under their believe that on a fundamental
level Nature was described by a local hidden variable (LHV) model,
i.e., that it is classical (realistic or deterministic) and local
(or non-signaling). The latter essentially means that no
information can travel faster than a maximal speed (e.g. of light)
which implies in particular that the probability distribution for
the outcomes of some experiment made by Alice cannot depend on
what other (spatially separated) physicist Bob does in his lab.
Otherwise, by choosing one or the other experiment, Bob could
influence instantly Alice's results and hence transmit information
at any speed. On the other hand, saying that Nature is classical
or deterministic means that the randomness in the outcomes that is
observed in the experiments comes from our ignorance of Nature,
instead of being an intrinsic property of it (as Quantum Mechanics
postulates). That is, Nature can stay in different {\it
configurations} $s$ with some probability $p(s)$ ($s$ is usually
called a hidden variable). But once it is in a fixed configuration
$s$, then any experiment has deterministic outputs. We note that
there are non-deterministic LHV models as well, but they can all
be cast into deterministic models \cite{WernerWolf}. Let us
formalize this a bit more.

\

Consider correlation experiments where each of ${ N}$ spatially
separated observers (Alice, Bob, Charlie,\ldots ) can measure
${ M}$ different observables with outcomes $\pm 1$: $\{A_{i_1}\}_{i_1=1}^M$ for Alice, $\{B_{i_2}\}_{i_2=1}^M$ for Bob and so on.
By repeating the experiment several times, for each possible
configuration of the observables (Alice measuring with the
aparatus $A_{i_1}$, Bob with the aparatus $B_{i_2}$, \ldots), they
can obtain a good approximation of the expected value of the
product of the outcomes of such configuration $\langle A_{i_1}
B_{i_2} C_{i_3}\cdots\rangle$. If Nature is described by a LHV
model, then
\begin{equation}\label{eq:classical-to-quantum}
\langle A_{i_1} B_{i_2} C_{i_3}\cdots\rangle=\langle A_{i_1}
B_{i_2} C_{i_3}\cdots\rangle_p= \sum_{s} p(s)
A_{i_1}(s)B_{i_2}(s)\cdots,\end{equation} where $A_{i_1}(s)=\pm 1$
is the deterministic outcome obtained by Alice if she does the
experiment $A_{i_1}$ and Nature is in state $s$ (notice that we
are including also the locality condition when assuming that
$A_{i_1}(s)$ is independent of $i_2,i_3,\ldots$).

For a quantum mechanical system in a state $\rho$ we have to set
\begin{equation}\label{eq:quantum-to-classical}
\langle A_{i_1} B_{i_2} C_{i_3}\cdots\rangle=\langle A_{i_1} B_{i_2}
C_{i_3}\cdots\rangle_\rho=\tr(\rho A_{i_1}\otimes B_{i_2}\otimes
C_{i_3}\cdots)\end{equation} where $\rho$ is a density operator
acting on a Hilbert space $\C^{d_1}\otimes\cdots\otimes \C^{d_{ N}}$
and the observables satisfy $-\1\leq
A_{i_1},B_{i_2},C_{i_3},\ldots\leq\1$, describing measurements
within the framework of positive operator valued measures (POVMs).
Note the parallelism with (\ref{eq:classical-to-quantum}). In fact
the quantum mechanical expression coincides with the classical one
if the matrices $A_{i_1}$'s, $B_{j_2}$'s, \ldots commute with each
other (and therefore can be taken diagonal in some basis $|s\>$),
and we take the state $\rho$ to be the separable state given by
$\rho=\sum_{s} p(s) |s\>\<s|\otimes |s\>\<s|\otimes \cdots$.

\

How can one then know if Nature allows for a LHV description or
follows Quantum Mechanics? That is, how to discriminate between
(\ref{eq:classical-to-quantum}) and
(\ref{eq:quantum-to-classical})? The key idea of Bell \cite{Bell}
was to realize that this can be done by taking linear combinations
of the expectation values $\langle A_{i_1} B_{i_2}
C_{i_3}\cdots\rangle$. So, given real coefficients $T_{i_1
i_2,\ldots}$, if we maximize the expression
\begin{equation}\label{eq:Bell-guay}
\left|\sum_{i_1,\cdots ,i_{ N}=0}^{ M-1} T_{i_1 \cdots i_{N}}\langle
A_{i_1} B_{i_2} C_{i_3}\cdots\rangle\right|\end{equation} assuming
(\ref{eq:classical-to-quantum}) we get \footnote{We write $\|T\|$
since the above expression is exactly the norm of $T$ as a
$N$-linear form from $\R^D$ equipped with the sup-norm.}
$$\|T\|:=\sup_{a_{i_1}, b_{i_2}, c_{i_3}, \ldots =\pm 1} \left|\sum_{i\in\mathbb{Z}_M^N} T_{i_1 \cdots i_{N}} a_{i_1}b_{i_2}c_{i_3}\cdots\right|.$$
Therefore, if all correlations predicted by quantum mechanics could
be explained in a classical and local world, one would have the
following {\it Bell inequality}:
\begin{equation}\label{eq:Bell-in}
\left|\sum_{i_1,i_2,i_3,\ldots} T_{i_1i_2i_3\ldots}\tr(\rho
A_{i_1}\otimes B_{i_2}\otimes C_{i_3}\cdots)\right|\le
\|T\|.\end{equation} However, Quantum Mechanics predicts examples
for which we have a {\it violation} in (\ref{eq:Bell-in}). The
largest possible violation of a given Bell inequality (specified by
$T$) within quantum mechanics is the smallest constant $K$ for which
\begin{equation}\label{eq:groth}
\left|\sum_{i_1,i_2,i_3,\ldots} T_{i_1i_2i_3\ldots}\tr(\rho
A_{i_1}\otimes B_{i_2}\otimes C_{i_3}\cdots)\right|\le
K\|T\|\end{equation} holds independent of the state and the
observables. For instance, for the CHSH inequality ($M=2$, $N=2$ and
$T$ the Hadamard matrix) we have $K=\sqrt{2}$ irrespective of the
Hilbert space dimension. More generally, if we also allow for
arbitrary ${ M}$ and $T$ and just fix ${ N}=2$, there is (see
Section \ref{sec:preliminaries}) a universal constant (called
Grothendieck's constant) $K_G$ that works in (\ref{eq:groth}) for
\textit{all} Bell inequalities, states and observables. This was
firstly observed by Tsirelson \cite{Tsirelson} (see also
\cite{Acin-groth}). As $K_G$ is known to lie in between $1.676..\leq
K_G \leq 1.782..$ the maximal Bell violation in Eq.(\ref{eq:groth})
is bounded for bipartite quantum systems. This bound imposes some
limitations to the use of Bell inequalities, where one usually
desires as large violation as possible. Below, when talking about
the implications of our main result, we will illustrate why having
large violations can be useful in the contexts of communication
complexity, quantum cryptography or noise robustness.

Therefore, it would be very useful to know whether in the tripartite
case we still have a uniform bound for the violations. The first
place in which we have found this question explicitely is in the
review \cite{Tsirelson} of Tsirelson in 1993. Our main result will
be to prove that this is not the case.

(We will in the following use $\succeq,\simeq,\preceq$ to denote
$\geq,=,\leq$ up to some universal constant).

\

\begin{thm}[Maximal violation for tripartite Bell inequalities]\label{mainBellThm}

\

\begin{enumerate}
\item For every dimension $d\in\mathbb{N}$, there exist
$D\in\mathbb{N}$, a pure state $|\psi\>$ on $\C^d\otimes \C^D
\otimes \C^D$ and a Bell inequality with traceless observables such
that the violation by $|\psi\>$ is $\succeq \sqrt{d}$.

\item The (unnormalized) state  can be taken $|\psi\>= \sum_{1\le
i\le d\; 1\le j,k\le D} \<j|U_{i}^\dagger|k\> \;|ijk\>$, where
$U_i$ are random unitaries.

\item The order ${\sqrt{d}}$ is optimal in the sense that,
conversely, for every state acting on $\C^d\otimes \C^D \otimes
\C^D$ and every Bell inequality with not necessarily traceless
observables the violation is also $\preceq \sqrt{d}$.
\end{enumerate}
\end{thm}

\begin{figure}[h]
 \includegraphics[width=12cm]{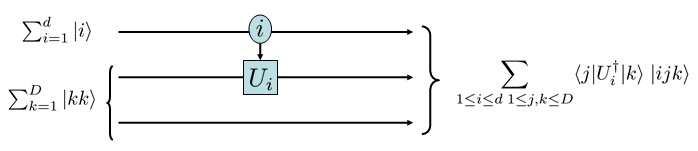}\\
 \caption{Quantum circuit that provides highly non-local states. Apart from using a maximally entangled state as an input,
 it requires the implementation of a {\it controlled unitary} with {\it random} unitaries (that is,
 if the control qubit is in state $|i\>$, the circuit applies the (random) unitary $U_i$)}\label{fig2}
\end{figure}

This Theorem shows once more that {\it random} states exhibit
unexpected extremal properties
\cite{random1,random2,random3,random4}. Unfortunately, though we
have a explicit form for these highly non-local states (see Figure
\ref{fig2}), there are a couple of weaknesses in the above theorem,
which mainly come from the techniques we use:

\begin{itemize}
\item We do not have any control on the growth of $D$ with respect
to $d$. \item We do not have a explicit form for the family of
inequalities for which we have unbounded violation. As it will be
shown in the proof, for both the choice of the observables and the
choice of the coefficients of the Bell inequality we will use a
{\it lifting} argument, which in our case goes back to some
application of Hahn-Banach's theorem and the clever use of
approximate unit in ideals of a $C^*$-algebra. This prevents us
from having a constructive proof.  It would be interesting to find
these lifting in another way (even numerically or
probabilistically).
\end{itemize}

It is important to note here (see Sec. \ref{sec:results2}) that in
contrast to what is known for the ${ M}=2$ case \cite{WW}, GHZ
states do not belong to this set of highly non-local states---they
always lead to a bounded violation. Let us now discuss some of the
implications of Thm.\ref{mainBellThm}:

\subparagraph{Communication complexity:}  Using notions from
\cite{CCP0} it was shown in \cite{CCP} that for every quantum state
that violates a Bell inequality there is a communication complexity
problem for which a protocol assisted by that state is more
efficient than any classical protocol. In fact, it turns out that
there is a quantitative relation between the amount of violation and
the superiority of the assisted protocol.

Adapted to our case, the communication complexity problem discussed
in \cite{CCP0,CCP,moresettings2} is the following: Each of the three
parties ($i=1,2,3$) obtains initially a random bit string encoding
$(x_i,y_i)$, where each $y_i=\pm 1$ is taken from a flat
distribution and $x_i\in\{0,\ldots,{ M-1}\}$ is distributed
according to $|T_x|/\sum_{x'}|T_{x'}|$ where $T_x=T_{x_1,x_2,x_3}$
are the coefficients appearing in the violated Bell inequality. The
goal is now that every party first broadcasts a single bit and then
attempts to compute the function
\begin{equation}
F(x,y)=\prod_iy_i T_x/|T_x|
\end{equation} upon the obtained information. The protocol was
successful if all parties come to the right conclusion. If one
compares the optimal classical protocol (assisted by shared
randomness) with a protocol assisted by a quantum state violating
the considered Bell inequality by a factor $K$, and denotes the
respective probabilities of success by $P$ and $P_K$ then
\begin{equation}
\frac{P_K-\frac12}{P-\frac12}=K.
\end{equation}
Let us denote by $H(P)$ the binary entropy and quantify the
information $I$ about the actual value of $F(x,y)$ gained by a
protocol with success probability $P$ by $I(P)=1-H(P)$. Taking the
states and inequalities appearing in Thm.\ref{mainBellThm} and thus
setting $K\succeq \sqrt{d}$ then leads to the ratio
\begin{equation}
\frac{I(P_K)}{I(P)}\succeq d\;.
\end{equation}

\subparagraph{Measuring the size of the Hilbert space:} What do
measured correlations tell us about a quantum system, if we do not
have a priori knowledge about the observables or even the size of
the underlying Hilbert space? This type of question becomes for
instance relevant in the context of cryptography where one wants
to avoid any kind of auxiliary assumption necessary for security
\cite{Acin0,Acin01,Acin02,Kent,MasanesWinter}. In the context of
detecting entanglement it is easy to see that the set of
entanglement witnesses that remain meaningful when disregarding
the Hilbert space dimension is exactly the set of Bell
inequalities. In fact, if measured correlations do not violate any
Bell inequality, then they can always be produced by a separable
(i.e., unentangled) state in a sufficiently large Hilbert space
\cite{Acin01}. Thm.\ref{mainBellThm} now shows that for
multipartite systems the violation of a Bell inequality can in
principle be used to estimate (lower bound) the Hilbert space
dimension. It also answers a question posed by Masanes
\cite{Masanes} in the negative: in contrast to the case ${ M}=2$
\cite{WW,Masanes} the extreme points of the set of quantum
correlations observable with dichotomic measurements are in
general not attained for multi-qubit systems.

\subparagraph{Robustness against noise and detector inefficiencies:}
It is well known that for ${ M}=2$ the maximal quantum violation can
increase exponentially in the number of sites $N$ \cite{MA,WW}.
However, since the $N$ parties have to measure in coincidence, in
practice with imperfect detectors, this increase comes with the
handicap that also the coincidence rates then decrease
exponentially. This becomes clearly different if one increases the
violation without increasing $N$ as it is the case in
Thm.\ref{mainBellThm}. So, in spite of the opaqueness of our result
concerning practical implementations it does not suffer from
decreasing coincidence rates.

Similarly, Thm.\ref{mainBellThm} implies the existence of tripartite
quantum states that can withstand an arbitrary amount of white noise
before they admit a description within a local hidden variable
model. To see this let $\rho$ belong to the family of states giving
rise to a maximal violation $K\succeq\sqrt{d}$ and set
\begin{equation}\label{eq:noise}\rho'=p\rho+(1-p)\frac{\1}{\tr(\1)}.\end{equation}
As the violation $K$ is attainable for traceless observables,
$\rho'$ yields $K'=p K$ which is still a violation whenever
$p\succeq1/\sqrt{d}$ (see \cite{Acin-groth} for a similar reasoning
in the bipartite case). In this context, it is a natural question to
ask which is the amount of noise needed to disentangle a quantum
state. It happens that this is considerably bigger. In particular,
it is shown in \cite{Rungta} (in a constructive way) that:

\begin{thm}[Neighborhood of the maximally mixed state]\label{thm:separable}
Given $d$, there is an entangled state $\rho_d$ in $\C^d\otimes \C^d
\otimes \C^d$ such that $\rho'_d=p\rho_d+ (1-p)\frac{\1}{d^3}$ is
still entangled whenever $p\succeq \frac{1}{d^2}$.
\end{thm}

We will give an independent proof in the Appendix. Up to now the
optimal value of $p$ is not known. The best bounds are given by
$\frac{1}{d^3}\preceq p\preceq \frac{1}{d^2}$ \cite{Rungta,
Gurvits}. It is also known that $\rho_d$ in Theorem
\ref{thm:separable} can be taken to be the generalized GHZ state
\cite{Deuar} in contrast to what we will see for the maximal
violation of multipartite Bell inequalities.


\section{Mathematical tools}\label{sec:preliminaries}

We will use tools from the theory of Operator Spaces and Tensor
Norms. The use of one or the other will depend on the point of
view of our problem. If we put the focus on the Bell inequalities
and ask for the largest possible violation within Quantum
Mechanics, then we will work with Operator Spaces and the {\it
meta-theorem} we have is the following (for a precise formulation
see below):

\

{\it A Bell inequality for $N$ observers and $M$ dichotomic
observables per site is given by a $N$-linear form
$T:\ell_\infty^M\times \cdots\times \ell_\infty^M\lra \C$ with
$\|T\|=1$. The largest possible violation within Quantum Mechanics
is given by the completely bounded norm of $T$, $\|T\|_{cb}$. }

\

If, however, we put the focus on the quantum states and ask, given a
$N$-partite quantum state, which is the largest possible violation
that this state gives to a Bell inequality, then we will work with
the theory of Tensor Norms, and the {\it meta-theorem} now reads:

\

{\it The largest possible violation that a $N$-partite $D\times
\cdots \times D$ state $\rho$ gives to a Bell inequality (with an
arbitrarily number of dichotomic observables) is given by the
extendible tensor norm
$$\|\rho\|_{\otimes^N_{j=1,\alpha_{ext}} S_1^D}.$$}

\subsection*{Operator spaces}

The theory of operator spaces started with the work of Effros and
Ruan in the 80's (see e.g. \cite{EffrosRuan,Pisierbook}) where
they characterized, in an abstract sense, the structure of a
subspace of a $C^*$-algebra. Since then, this theory has found
some interesting applications in mathematical analysis. An
operator space is a complex vector space $E$ and a sequence of
norms $\|\cdot\|_n$ in the space of $E$-valued matrices
$M_n(E)=M_n\otimes E$, which verify the properties
\begin{enumerate}
\item For all $n$, $x\in M_n(E)$ and $a,b\in M_n$ we have that
$\|axb\|_n\le \|a\|_{M_n}\|x\|_n\|b\|_{M_n}$. \item For all $n,m$,
$x\in M_n(E)$, $y\in M_m(E)$, we have that $$\left\|\left(%
\begin{array}{cc}
 x & 0 \\
 0 & y \\
\end{array}%
\right)\right\|_{n+m}= \max\{\|x\|_n,\|y\|_m\}$$
\end{enumerate}

Any $C^*$-algebra has a natural operator space structure that is the
resulting of embedding it inside the space $\mathcal{B}(H)$ of
bounded linear operators in a Hilbert space \cite{EffrosRuan,
Pisierbook}, where $M_n(\B(H))=\B(\ell_2^n\otimes H)$. In
particular, $\ell_{\infty}^k$ ($=\C^k$ with the sup-norm), being a
commutative $C^*$-algebra, has a natural operator space structure.
To compute it we embed $\ell_{\infty}^k$ in the diagonal of $M_k$
(with the operator norm) and then, given $x=\sum_i A_i \otimes
|i\>\in M_n(\ell_{\infty}^k)=M_n\otimes \ell_{\infty}^k$, we have
\begin{equation}\label{eq:op-space-infty}
\|x\|_n=\left\|\sum_{i} A_i \otimes
|i\>\<i|\right\|_{M_{nk}}=\max_i\|A_i\|_{M_n}.
\end{equation}

The morphisms in the category of operator spaces (that is, the
operations that preserve the structure) are called completely
bounded maps. They are linear maps $u:E\lra F$ between operator
spaces such that all the amplifications $u_n=\1_n \otimes u
:M_n(E)\lra M_n(F)$ are bounded. The cb-norm of $u$ is then
defined as $\|u\|_{cb}=\sup_n\|u_n\|$. We will call $CB(E,F)$ the
resulting normed space, that is, in fact, an operator space by
$M_n(CB(E,F))=CB(E, M_n(F))$. Analogously one can define the
cb-norm of a multilinear map $T:E_1\times\cdots\times E_N\lra F$
as $\|T\|_{cb}=\sup\|T_{n_1,\ldots,n_N}\|$, where now
$T_{n_1,\ldots,n_N}=T\otimes \1_{n_1}\otimes \cdots\otimes
\1_{n_N}:M_{n_1}(E_1)\times\cdots\times M_{n_N}(E_N)\lra
M_{n_1\cdots n_N}(F)$. A multilinear map is called completely
bounded if $\|T\|_{cb}<\infty$. We will denote by
$CB^N(E_1,\ldots,E_N;F)$  the resulting normed space, that is also
an operator space by
$M_n(CB^N(E_1,\ldots,E_N;F))=CB^N(E_1,\ldots,E_N;M_n(F))$.

\

With these definitions, if we have a $N$-linear form
$T:\ell_\infty^M\times \cdots \times \ell_\infty^M\lra \C$ given by
$T(|i_1\>,|i_2\>,\ldots)=T_{i_1 i_2 \ldots}$ and we compute the
usual norm and the cb-norm we obtain
\begin{align*}
\|T\|&=\sup\{\left|\sum_{i_1,\cdots , i_{ N}=0}^{M-1} T_{i_1 i_2
\ldots}\epsilon^1_{i_1}
\epsilon^2_{i_2}\cdots\right|\; ; |\epsilon^j_{i_j}|\le 1\},\\
\|T\|_{cb}&=\sup\left\{\left|\sum_{i_1,\cdots , i_{ N}=0}^{M-1}
T_{i_1 i_2 \ldots}\tr(\rho A_{i_1}\otimes B_{i_2}\cdots)\right|\; :
\;
\begin{array}{c}
 A_{i_1}, B_{i_2}, \ldots \in M_D \text{ with operator norm } \le
1 \\
\rho \in M_{D^N} \text{ with trace norm } \le 1 \\
\end{array} \; \right\}.
\end{align*}
These expressions coincide respectively with the maximal value that
one can achieve in the expression (\ref{eq:Bell-guay})
$$\left|\sum_{i_1,\cdots ,i_{
N}=0}^{ M-1} T_{i_1 \cdots i_{N}}\langle A_{i_1} B_{i_2}
C_{i_3}\cdots\rangle\right|$$ if we assume that Nature is
deterministic and local, $\|T\|$, or if we assume Quantum Mechanics,
$\|T\|_{cb}$. This essentially proves the {\it meta theorem} stated
at the beginning of the Section. The only subtle point is that in
the context of Bell inequalities everything is {\it real} while in
this context of operator spaces we are in the {\it complex} case.
Therefore, whenever we want to formally use this {\it meta-theorem}
we will have to make some splits between real and imaginary parts.

\

In \cite{Grothendieck}, Grothendieck proved what he called the {\it
fundamental theorem of the metric theory of tensor products}. This
result, known as Grothendieck's Theorem or Grothendieck's Inequality
reads as follows:

\

{\it There exists a universal constant $K_G$ such that no matter how
we choose real coefficients $T_{ij}$ and  elements $x_i, y_j$ in the
unit ball of a real Hilbet space $H$ with inner product
$\<\cdot,\cdot\>$, we have that
$$\left|\sum_{ij}T_{ij}\<x_i,y_j\>\right|\le K_G \sup_{\epsilon_i,\nu_j=\pm
1}\left|\sum_{ij}T_{ij}\epsilon_i\nu_j\right|.$$ In particular,
\begin{equation}\label{KGv}
\left|\sum_{ij}T_{ij}tr(\rho A_i\ten B_j)\right|\le K_G
\sup_i\|A_i\| \sup_j \|B_j\| \sup_{\epsilon_i,\nu_j=\pm
1}\left|\sum_{ij}T_{ij}\epsilon_i\nu_j\right|. \end{equation} }

The second part tells us that Grothendieck's Theorem provides a
 uniform bound $K_G$ for the violation of any bipartite Bell
inequality with dichotomic observables. This was essentially
Tsirelson's observation \cite{Tsirelson}.

\

But the above comments show how Grothendieck's Theorem also implies
that any bounded bilinear form from a commutative $C^*$-algebra has
to be also completely bounded, which was firstly noticed in
\cite{KumarSinclair}.

\

Since Grothendieck stated his Theorem, a lot of effort has been
devoted to find suitable multilinear generalizations (see for
instance
\cite{groth-mult1,groth-mult2,groth-mult3,groth-mult4,groth-mult5,perez-q-algebra}).
However, up to know, the validity of a trilinear Grothendieck's
Theorem in the context of operator spaces (and hence in the
context of Bell Inequalities) has been open. Although it is
conceivable that trilinear versions of Grothendieck's inequality
hold for operator spaces,  our main theorem (Theorem
\ref{mainBellThm}) will show that the trilinear version of
\eqref{KGv} fails. We will rewrite this now in the language of
operator spaces which is instrumental in the proof. Then we will
show how this Theorem implies Theorem \ref{mainBellThm} and we
will give the proof in Section \ref{sec:proofs}.

\begin{thm}\label{thm:marius}
For every $n$, there exist $N$, a state $|\psi_N\>$, a trilinear
form $T:\ell_{\infty}^{2^{n^2}}\times \ell_\infty^{2^{N^2}} \times
\ell_\infty^{2^{N^2}}\lra \C$ and elements $b\in
M_n(\ell_\infty^{2^{n^2}})$, $\hat{b}\in
M_n(\ell_\infty^{2^{N^2}})$, with  $\||\psi_N\>\|,\|T\|, \|b\|,
\|\hat{b}\|\preceq 1$ and
$$\left|\<\psi_N|T_{n,N,N}(b,\hat{b},\hat{b})|\psi_N\>\right|\succeq \sqrt{n}.$$
Moreover
\begin{enumerate}
\item The order $\sqrt{n}$ is optimal. \item $|\psi_N\>$ can be
taken $\frac{1}{\sqrt{nN}} \sum_{1\le i\le n\; 1\le j,k\le N}
\<j|U_{N,i}^\dagger|k\> |ijk\>$ where $U_{N,i}$ are random
unitaries.
\end{enumerate}
\end{thm}

\noindent Again we use $\succeq$ (resp. $\simeq$) to denote $\ge$
(resp. $=$) up to some universal constant. In particular, we obtain
that

\begin{cor}[Bounded but not completely bounded trilinear forms]
Given $n$, there exists $N$ and a trilinear map
$T:\ell_\infty^n\times \ell_\infty^N\times \ell_\infty^N\lra \C$
such that $\|T\|_{cb}\ge \|T_{n,N,N}\|\succeq\sqrt{n}\|T\|$.
Moreover, the order $\sqrt{n}$ is optimal.
\end{cor}

\


We will finish this section by showing how Theorem \ref{thm:marius}
implies Theorem \ref{mainBellThm}. As we said before, it is simply a
matter of splitting into real and imaginary parts.

\

Theorem \ref{thm:marius} tells us that there exist a complex matrix
$\{T_{ijk}\}_{i,j,k=1}^N$, $n\times n$ matrices $b_i$ and $N\times
N$  matrices $\hat{b}_j$ (all of them with norm $\preceq 1$) such
that
\begin{equation}\label{eq:groth2}
\left|\<\psi_N|\sum_{i,j,k} T_{ijk} b_i\otimes \hat{b}_j\otimes
\hat{b}_k|\psi_N\>\right|\succeq \sqrt{n}.
\end{equation}

By splitting into real and imaginary parts it is not difficult to
see that one can take in (\ref{eq:groth2}) $T$ real and $b_i,
\hat{b}_j$ hermitian. Moreover, writing
$\alpha_i=\frac{\tr(b_i)}{n}$ (so that $|\alpha_i|\le 1$),
$b^1_i=\alpha_i \1_n$, $b^0_i= b_i-b^1_i$ and applying the bipartite
case to show that ($\rho=|\psi_N\>\<\psi_N|$)
$$\left|\sum_{i,j,k} \tr(\rho T_{ijk} b^1_i\otimes
\hat{b}_j\otimes \hat{b}_k)\right|= \left|\sum_{i,j,k}
T_{ijk}\alpha_i \tr(\rho_{2,3} \hat{b}_j\otimes \hat{b}_k)\right|\le
K_G,$$ one can take the observables in (\ref{eq:groth2}) to be
traceless.


\subsection*{Tensor norms}

The theory of tensor norms can be traced back to the work of Murray
and von Neumann in the late 30's, but it was definitely set by
Grothendieck in his seminal paper \cite{Grothendieck}. Since then,
several and important contributions have been made (see
\cite{Defant} for a modern reference).

If $X_1,\ldots,X_N$ are normed spaces, by $\bigotimes_{j=1,\pi}^N
X_j$ we denote the algebraic tensor product $\bigotimes_{j=1}^N X_j$
endowed with the projective norm
$$\pi(u):=\inf\left\{\sum_{i=1}^m \|u^1_i\|\cdots\|u^N_i\|:u=\sum_{i=1}^mu^1_i\otimes\cdots\otimes u^N_i\right\}.$$
This tensor norm is both commutative and associative, in the sense
that $\bigotimes_{j=1,\pi}^N X_j=\bigotimes_{j=1,\pi}^N
X_{\sigma(j)}$ for any permutation of the indices $\sigma$ and that
$\bigotimes_{j=1,\pi}^N \left(\bigotimes_{i_j=1,\pi}^{N_j}
X^j_{i_j}\right)=\bigotimes_{j=1,i_j=1,\pi}^{N,N_j} X^j_{i_j}$. The
projective norm $\pi$ is in duality with the injective norm
$\epsilon$, defined on $\bigotimes_{j=1}^N X_j$ as
$$\epsilon(u):=\sup\left\{\left|\sum_{i=1}^m \phi^1(u^1_i)\cdots \phi^N(u^N_i)\right|:
\phi^j\in X_j^*, \|\phi^j\|\le 1\right\}\; ,$$ where $X_j^*$ denotes
the topological dual of $X_j$ and $u=\sum_{i=1}^m u^1_i\otimes
\cdots\otimes u^N_i$. That is, if $E_j$ is a finite dimensional
normed space for every $j=1,\cdots ,N$, we have
$\left(\bigotimes_{j=1,\pi}^N
E_j\right)^*=\bigotimes_{j=1,\epsilon}^N E^*_j$. Moreover, the dual
of the $\pi$ tensor product can also be isometrically identified
with the space of $N$-linear forms (with its usual operator norm).
In fact, we have the natural isometric identification,
\begin{equation}\label{eq:duality-pi-epsilon}
\mathcal{L}^N(E_1,\ldots, E_N; \C)=(\bigotimes_{j=1,\pi}^N
E_j)^*=\mathcal{L}(E_1,\bigotimes_{j=2,\epsilon}^N E^*_j).
\end{equation}

Following \cite{Defant} (or \cite{Floret} for the multilinear
version) we define a tensor norm $\beta$ of order $N$ as a way of
assigning to every $N$-tuple of normed spaces $(X_1,\ldots, X_N)$ a
norm on $\bigotimes_{j=1}^NX_j$ (we call $\bigotimes_{j=1,
\beta}^NX_j$ to the resulting normed space) such that

\begin{itemize}
\item $\epsilon\le \beta\le \pi$ \item $\left\|\bigotimes_{j=1}^N
u_j:\bigotimes_{j=1, \beta}^NX_j\lra \bigotimes_{j=1,
\beta}^NY_j\right\|\le \prod_{j=1}^N\|u_j\|$, for every choice of
linear bounded operators $u_j:X_j\lra Y_j$. This is called the {\it
metric mapping property}.
\end{itemize}

Sometimes we will use the notation $\bigotimes_\beta(X_1,\cdots
,X_N)$ instead of $\bigotimes_{j=1, \beta}^NX_j$ to distinguish
some space.

We will say that $\beta$ is finitely generated if, for every $X_j$,
$j=1,\cdots ,N$, and $z\in \bigotimes_{j=1}^NX_j$ we have
$$\beta(z;X_1,\cdots ,X_N)=\inf\{\beta(z;E_1,\cdots ,E_N):
E_j\in FIN(X_j), z\in \bigotimes_{j=1}^N E_j\};$$ where we denote
$FIN(X)=\{E\subset X| \dim E<\infty\}$ (and $COFIN(X)=\{E\subset X|
E \text{ is closed and } \dim(X/E)<\infty\}$).

As one can find in \cite[Sec. 17]{Defant} and \cite[Sec.
4]{Floret} tensor norms are in one-to-one duality with ideals of
multilinear operators. We explain this in which follows:

A normed (Banach) ideal of $N$-linear continuous operators between
Banach spaces is a pair $(\mathcal{A},\|\cdot\|_{\mathcal{A}})$ such
that

\begin{itemize}
\item $\mathcal{A}(X_1,\cdots,X_N;Y)=\mathcal{A}\cap
\mathcal{L}^N(X_1,\cdots,X_N;Y)$ is a linear subspace of
$\mathcal{L}^N(X_1,\cdots,X_N;Y)$ and the restriction
$\|\cdot\|_{\mathcal{A}}|_{\mathcal{A}(X_1,\cdots,X_N;Y)}$ is a
(complete) norm.

\item If $u_j\in \mathcal{L}(Z_j,X_j)$, $T\in
\mathcal{A}(X_1,\cdots,X_N;Y)$ and $v\in \mathcal{L}(Y,Z)$, then the
composition $v\circ T\circ (u_1,\cdots ,u_N)$ is in $\mathcal{A}$,
and $$\|v\circ T\circ (u_1,\cdots ,u_N)\|_{\mathcal{A}}\leq
\|v\|\|T\|_{\mathcal{A}}\|u_1\|\cdots \|u_N\|.$$

\item The operator $\mathbb{K}^N\ni (x_1,\cdots ,x_N)\mapsto
x_1\cdots x_N\in \mathbb{K}$ is in $\mathcal{A}$ and it has
$\|\cdot\|_{\mathcal{A}}$-norm equal to one.
\end{itemize}

An ideal $(\mathcal{A},\|\cdot\|_{\mathcal{A}})$ is called maximal
if $\|T\|_{\mathcal{A}^{\max}}:=\sup \{\|q_L^Y \circ T|_{E_1\times
\cdots \times E_N}\|_\mathcal{A}| E_j\in FIN(X_j), L\in
COFIN(Y)\}<\infty$ implies $T\in \mathcal{A}$ and
$\|T\|_{\mathcal{A}}=\|T\|_{\mathcal{A}^{\max}}$.

The following theorem shows the duality mentioned above

\begin{thm}\label{tensor-ideal}
Let $(\mathcal{A},\|\cdot\|_{\mathcal{A}})$ be a normed ideal of
N-linear continuous mappings between Banach spaces. Then
$(\mathcal{A},\|\cdot\|_{\mathcal{A}})$ is maximal if and only if
there exists a finitely generated tensor norm $\beta$ of order $N+1$
such that

$$\mathcal{A}(X_1,\cdots,X_N;Y^*)=(\bigotimes_\beta(X_1,\cdots
,X_N,Y))^*,$$

$$\mathcal{A}(X_1,\cdots,X_N;Y)=(\bigotimes_\beta(X_1,\cdots
,X_N,Y^*))^*\cap \mathcal{L}^N(X_1,\cdots,X_N;Y).$$

Here both identifications are isometric.

\end{thm}

For the purposes of this paper we will only need two of these
ideals: the extendible and the $(1;2)$-summing multilinear
operators.

\subsubsection*{Extendible multilinear operators}

The lack of a multilinear Hahn-Banach extension theorem has
motivated a considerable effort in the search of partial positive
results (see \cite{Hahn-Banach1,Hahn-Banach2,Hahn-Banach3,
Hans-paper} and the references therein). In this context, the
natural space to work with is the space of {\it extendible}
multilinear forms. That is, those continuous multilinear forms
$T:X_1\times \cdots \times X_n\lra \C$ (here $X_j$ are Banach
spaces) such that for every choice of superspaces $Y_j\supset X_j$,
there is a continuous and multilinear extension $\tilde{T}:Y_1\times
\cdots \times Y_n\lra \C$. We define the {\it extendible} norm of
$T$ as
$$\|T\|_{ext}=\sup_{Y_j}\inf_{\tilde{T}}\|\tilde{T}\|,$$
where the sup runs among all possible superspaces $Y_j$ and the
inf among all possible extensions $\tilde{T}$. As it can be found
in \cite{Hahn-Banach2}, for infinite dimensional spaces $X_j$,
$\|T\|_{ext}$ can be in general $\infty$. We say then that $T$ is
extendible if $\|T\|_{ext}<\infty$.

It can be easily seen that the extendible n-linear forms constitute
a Banach ideal, which we denote by $\mathcal{L}^n_{ext}$. Actually,
it is trivial to check that $(\bigotimes_{j=1,
\alpha_{ext}}^nX_j)^*=\mathcal{L}^n_{ext}(X_1,\cdots ,X_n)$
isometrically, if we define the well known (\cite{KiRa}, Sec. 3)
finitely generated tensor norm

$$\alpha_{ext}(u;X_1,\cdots ,X_n)=\inf\{\pi(u; Y_1,\cdots ,Y_n):
X_j\subset Y_j\},$$ where the $\inf$ is taken among all superspaces
$Y_j$ of $X_j$.

$\alpha_{ext}$ is called the extendible tensor norm (and it is, of
course, the tensor norm associated to the ideal of extendible
multilinear forms in the sense of Theorem \ref{tensor-ideal}).

The next lemma will be a central result to connect this mathematical
theory with the context of Bell inequalities:

\begin{lem}
Let $X_1,\cdots ,X_n$ be $n$ Banach spaces and $u\in
\bigotimes_{j=1}^nX_j$. We have
$$\alpha_{ext}(u)=\sup_{M,A^1_{i_1},\cdots ,A^n_{i_n}}\left|\sum_{i_1,\cdots,i_n=1}^k M_{i_1\cdots i_n}
\langle A^1_{i_1}\otimes\cdots \otimes
A^n_{i_n},u\rangle\right|,$$ where the $\sup$ is taken among
$(A^1_{i_1})_{i_1=1}^k\subset B_{{X^*_1}},\cdots,
(A^n_{i_n})_{i_n=1}^k\subset B_{{X^*_n}},$ $\|(M_{i_1\cdots
i_n})_{i_1,\cdots
,i_n=1}^k\|_{\bigotimes_{j=1,\varepsilon}^n\ell_1^k}\leq1$, $k\in
\mathbb{N}$, and the brackets denote as usual the action by
duality.

\end{lem}

\begin{proof}

By the injectivity of $\ell_\infty$ (see for instance \cite[Chap
I.1]{Defant}), it follows that
$$\alpha_{ext}(u)=\sup \left\{\|a_1\otimes \cdots \otimes
a_n(u)\|_{\bigotimes_{j=1,\pi}^n\ell_\infty^k}\mid
a_j:X_j\rightarrow\ell_\infty^k,
\|a_j\|\le 1, j=1,\cdots,n; k\in \mathbb{N} \right\}.$$

Now, we know that $\mathcal{L}(X_j,\ell_\infty^k)$ is isometrically
isomorphic to $\ell_\infty^k(X^*_j)$ (see for instance \cite[Chap
I.3]{Defant}). Thus, given $a_j\in
\mathcal{L}(X_j,\ell_\infty^k)=\ell_\infty^k(X^*_j)$ by
$a_j=\sum_{i_j=1}^k |i_j\rangle \otimes A_{i_j}$, we have
$$\|a_1\otimes \cdots \otimes
a_n(u)\|_{\bigotimes_{j=1,\pi}^n\ell_\infty^k}=\|\sum_{i_1,\cdots,i_n=1}^k\langle
A^1_{i_1}\otimes\cdots \otimes A^n_{i_n},u \rangle
|i_1\>\otimes\cdots \otimes
|i_n\>\|_{\bigotimes_{j=1,\pi}^n\ell_\infty^k}=$$ $$ =\sup
\{|\sum_{i_1,\cdots,i_n=1}^k\langle A^1_{i_1}\otimes\cdots \otimes
A^n_{i_n},u \rangle T(|i_1\>,\cdots, |i_n\>)|: T\in
B_{(\bigotimes_{j=1,\pi}^n\ell_\infty^k)^*}\}.$$

The statement follows now easily.
\end{proof}

With this at hand we can now formalize the {\it meta-theorem} given
at the introduction of Section \ref{sec:preliminaries}:

\begin{thm}\label{thm:state-tensor-norm}
Given a $N$-partite $D_1\times \cdots \times D_N$ quantum state
$\rho$, the largest possible violation that this state gives to a
Bell inequality of an arbitrarily number of dichotomic observables
is upper bounded by $$2^{N-1}
\|\rho\|_{\bigotimes_{j=1,\alpha_{ext}}^NS_1^{D_j}}.$$
\end{thm}

\begin{proof}
Given a Bell inequality with (real) coefficients
$T_{i_1,\ldots,i_N}$ and observables $-\1\le A_{i_1}, B_{i_2},
\ldots \le \1$, it is clear by the definition of $\alpha_{ext}$ that
$$\left|\sum_{i_1,i_2,\ldots} T_{i_1,i_2,\ldots} \tr(\rho A_{i_1}\otimes B_{i_2}\otimes \cdots)\right|\le \|\rho\|_{\bigotimes_{j=1,\alpha_{ext}}^NS_1^{D_j}}
\sup_{\epsilon^j_{i_j}\in \C, |\epsilon^j_{i_j}|=1}
\left|\sum_{i_1,i_2,\ldots} T_{i_1,i_2,\ldots}
\epsilon^1_{i_1}\epsilon^2_{i_2}\cdots\right|.$$

To finish the proof of the Theorem it is enough to notice that (see
\cite[Proposition 19]{Gustavo})
$$\sup_{\epsilon^j_{i_j}\in \C,
|\epsilon^j_{i_j}|=1} \left|\sum_{i_1,i_2,\ldots} T_{i_1,i_2,\ldots}
\epsilon^1_{i_1}\epsilon^2_{i_2}\cdots\right|\le 2^{N-1}
\sup_{\epsilon^j_{i_j}=\pm 1} \left|\sum_{i_1,i_2,\ldots}
T_{i_1,i_2,\ldots} \epsilon^1_{i_1}\epsilon^2_{i_2}\cdots\right|.
$$

\end{proof}

\subsubsection*{Summing operators}
Since the work of Grothendieck \cite{Grothendieck}, the class of
absolutely summing linear operators plays a crucial role in the
theory of tensor norms (see \cite{DiJaTo} for a reference).
Motivated by that, A. Pietsch defined in \cite{Pietsch} the
following class of multilinear operators:

A multilinear form $T:X_1\times \cdots \times X_N\lra \C$ is called
$(s;r)$-summing $(1\le s,r<\infty)$ if there exists a constant $K$
such that for any choice of finite sequences $(x^j_i)_i\subset X_j$,
we have that
\begin{equation}\label{eq:summing}
\left(\sum_i\left|T(x^1_i,\ldots,x^N_i)\right|^s\right)^{\frac{1}{s}}\le
K \prod_{j=1}^N \|(x^j_i)_i\|_{r}^{\omega},
\end{equation}
where $\|(x^j_i)_i\|_{r}^{\omega}$ denotes the supremum, among all
elements $x_j^*$ in the unit ball of the dual space $X_j^*$, of
$$\left(\sum_i
\left|x^*_j(x^j_i)\right|^{r}\right)^{\frac{1}{r}}.$$ The smallest
$K$ valid in equation (\ref{eq:summing}) is called the $(s;r)$
norm of $T$, and we write $\|T\|_{(s;r)}$. The key result is the
following generalization of Grothendieck's inequality, which
appears explicitly in \cite[Corollary 2.5]{perez-q-algebra} (see
also \cite{groth-mult1,groth-mult2,Hans-paper,groth-mult3}).

\begin{thm}\label{Thm:Tsumming}
Every extendible $N$-linear form $T$ is $(1;2)$-summing and
$\|T\|_{(1;2)}\le K_G2^{\frac{N-2}{2}}\|T\|_{ext}$, where $K_G$ is
Grothendieck's constant.
\end{thm}


\section{Bounded violations for GHZ states}\label{sec:results2}

The maximal violation of  multipartite Bell inequalities with two
dichotomic observables per site \cite{WW,ZB,MA} is known to be
attained for  GHZ states
$|\psi\rangle=\frac{1}{\sqrt{n}}\sum_{i=0}^{n-1}|iii\rangle$ (where
$n=2$ is sufficient in this case). In contrast to that, we will show
here that GHZ states do not give rise to the maximal violation in
Thm.\ref{mainBellThm} but rather lead to a bounded violation. In
other words, there is a fixed amount of noise (independent of the
dimension) which makes the considered correlations of the GHZ state
admit a description within a local hidden variable model.

\

Before proving that we need a bit of work. We call $\rho$  the
unnormalized GHZ state $\sum_{ij} |i\>\<j|\otimes |i\>\<j|\otimes
|i\>\<j|$ as a member of $\bigotimes_{j=1}^3S_1^n$ ($S_1^n$ the
Banach space of trace class operators on a $n$-dimensional Hilbert
space $\ell_2^n$). When we consider a tensor norm $\alpha$ on
$\bigotimes_{j=1}^3S_1^n$, $\rho^*$ will be the same element as
$\rho$, but considered in the dual
$(\bigotimes_{j=1,\alpha}^3S_1^n)^*$. The key point is the following
result,

\begin{prop}\label{ideals} For every tensor norm $\alpha$,
$$\|\rho\|_{\bigotimes_{j=1,\alpha}^3 S_1^n}\cdot\|\rho^*\|_{(\bigotimes_{j=1,\alpha}^3 S_1^n)^*}=n^2.$$
\end{prop}

We will follow \cite[Theorem 2.5]{GL}. First we will need the next

\begin{lem}\label{tec}
Let $\alpha$ be any tensor norm and $A=\otimes_{j=1,\alpha}^3
S_1^n$. Let $G$ be a topological compact group such that $G\subset
isom(A,A)$, the group of isometries of $A$. We suppose:
\begin{enumerate}
\item[(i)] $g \rho=\rho$ for every $g\in G$. \item[(ii)] Given
$L\in A^*$, if $L\circ g=L$ for every $g\in G$, then $L=\lambda
\rho^*$ for some constant $\lambda$.
\end{enumerate}
Then we have that $\|\rho\|_{A}\cdot \|\rho^*\|_{A^*}=n^2$.
\end{lem}

\begin{proof}

Let us take $L\in A^*$ such that $\|L\|_{A^*}$=1 and
$L(\rho)=\|\rho\|_A$. Let $dg$ be the Haar measure on $G$. We define
$L_0=\int_{G}L\circ g dg$. It is easy to see that $L_0$ is well
defined and belongs to $A^*$ with $\|L_0\|_{A^*}\leq \|L\|_{A^*}$.
Now, by (i), $L_0(\rho)=\int_{G}L\circ g(\rho) dg=L(\rho)$.

On the other hand, for every $g'\in G$ we have $L_0\circ
g'=\int_{G}L\circ g\circ g' dg=\int_{G}L\circ gdg=L_0$, where we
have used the translational invariance of the Haar measure. Using
(ii) we conclude that $L_0=\lambda \rho^*$. We have
$$\|\rho\|_A=L(\rho)=L_0(\rho)=\lambda \rho^*(\rho)=\lambda n^2.$$
And also
$$\lambda\|\rho^*\|_{A^*}=\|\lambda\rho^*\|_{A^*}=\|L_0\|_{A^*}\leq
\|L\|_{A^*}=1.$$ Then $\|\rho^*\|_{A^*}\leq\frac{1}{\lambda},$ and
thus $\|\rho\|_{A}\cdot \|\rho^*\|_{A^*}\geq n^2$. The other
inequality is trivial.

\end{proof}

Using the previous lemma we can easily prove Proposition
\ref{ideals}:

\begin{proof}
We only need to show that there exists a topological compact
subgroup of $isom(A,A)$ which verifies the hypothesis of Lemma
\ref{tec}.

\

For every $\varepsilon=(\epsilon_1,\cdots ,\epsilon_n)$, where
$\epsilon_i=\pm 1$, we consider $g_{\varepsilon}:\mathbb{C}^n\lra
\mathbb{C}^n$ such that $g_{\varepsilon}(|i\>)=\epsilon_i |i\>$. For
every $\sigma$ permutation of $\{1,\cdots ,n\}$ we consider
$h_{\sigma}:\mathbb{C}^n\lra \mathbb{C}^n$ such that
$h_{\sigma}(|i\>)=|\sigma(i)\>$. Now we take the group ${G}$
generated by the elements of the form
$$(g_{\varepsilon}^*\otimes g_{\theta}) \otimes
(g_{\varepsilon}^*\otimes g_{\theta})\otimes id\,,\;
(g_{\varepsilon}^*\otimes g_{\theta}) \otimes id\otimes
(g_{\varepsilon}^*\otimes g_{\theta}) \text{ and }
(h_\sigma^*\otimes h_\tau)\otimes(h_\sigma^*\otimes
h_\tau)\otimes(h_\sigma^*\otimes h_\tau).$$

It is clear that $G$ is a compact subgroup of $isom(A,A)$ and that
${G}$ verifies (i). Let us check (ii). Let
$$L=\sum_{i,j,k,l,m,n}\lambda_{i,j,k,l,m,n}|i\>\<j|\otimes
|k\>\<l|\otimes |m\>\<n|$$ be an arbitrary element of $A^*$. We take
${g}=(g_{\varepsilon}^*\otimes g_{\theta}) \otimes id \otimes
(g_{\varepsilon}^*\otimes g_{\theta})$ in ${G}$. If we have
$L=L\circ {g}$, we get, for every $i,j,k,l,m,n$,
$$\lambda_{i,j,k,l,m,n}=\lambda_{i,j,k,l,m,n}\epsilon_i\theta_j\epsilon_m\theta_n,$$
for every choice of signs $\epsilon_i, \theta_j$. Therefore,
$\lambda_{i,j,k,l,m,n}=\lambda_{i,j,k,l,m,n}\delta_{i,m}\delta_{j,n}$.
Then
$$L=\sum_{i,j,k,l}\lambda_{i,j,k,l}|i\>\<j|\otimes |k\>\<l|\otimes
|i\>\<j|.$$ We can repeat the step before (taking now
$g=(g_{\varepsilon}^*\otimes g_{\theta}) \otimes
(g_{\varepsilon}^*\otimes g_{\theta})\otimes id$ to see that, in
fact,
$$L=\sum_{i,j}\lambda_{i,j}|i\>\<j|\otimes |i\>\<j|\otimes
|i\>\<j|.$$

Finally, taking $g=(h_\sigma^*\otimes
h_\tau)\otimes(h_\sigma^*\otimes h_\tau)\otimes(h_\sigma^*\otimes
h_\tau)\in {G},$ we see that
$\lambda_{i,j}=\lambda_{\sigma(i),\tau(j)}$ for every permutations
$\tau, \sigma$ and every $i,j$. Then, we get that
$\lambda_{i,j}=\lambda$, which finishes the proof.
\end{proof}

And finally we can get the desired bound for the GHZ violation:

\begin{thm}[GHZ bound]\label{Main}
Given the tripartite GHZ state
$|\psi\>=\frac{1}{\sqrt{n}}\sum_{i=0}^{n-1}|iii\>$, the largest
possible quantum violation for a Bell inequality with dichotomic
observables is upper bounded by $4\sqrt{2}K_G$.
\end{thm}
\begin{proof}
By Theorem \ref{thm:state-tensor-norm} it is enough to show that for
the unnormalized GHZ state $\rho=\sum_{i,j=0}^{n-1} |i\>\<j|\otimes
|i\>\<j|\otimes |i\>\<j|$ we have
$\|\rho\|_{\bigotimes_{j=1,\alpha_{ext}}^3S_1^n}\leq K\cdot n.$ Due
to Proposition \ref{ideals} we only have to prove that
$\|\rho^*\|_{ext}\geq \frac{1}{K}n$. To see this, we use that by
Thm.\ref{Thm:Tsumming} $\|T\|_{(1;2)}\leq K \|T\|_{ext},$ and then
it remains to be proven that $\|\rho^*\|_{{\mathcal
L}_{(1;2)}^3(S_1^n)}\geq n$. For that we consider the sequence
$(|r\>\<0|)_{r=1}^n\subset S_1^n$, which verifies
$\|(|r\>\<0|)_r\|_2^w\leq 1$, and $\sum_r \rho^*(|r\>\<0|,
|r\>\<0|,|r\>\<0|)=n$.
\end{proof}

\begin{obs}
Note that Theorem \ref{Main} holds also for $N$ parties, where now
the constant can be taken $K_G (2\sqrt{2})^{N-1}$. In
\cite{Zukowski93} an explicit set of inequalities was derived for
which GHZ states achieve a violation of the order $(\pi/2)^N$.
\end{obs}

\section{Proof of the Main Theorem} \label{sec:proofs}

\subsection*{More on operator spaces}

Ruan's Theorem \cite{Pisierbook,EffrosRuan,Ruan} assures that any
operator space can be considered as a closed subspace of
$\mathcal{B}(H)$ with the inherited sequence of matrix norms. Then
we can define the minimal tensor product of two operator spaces
$E\subset \B(H)$ and $F\subset \B(K)$ as the operator space given by
$$E\otimes_{\min} F\subset \B(H\otimes K). $$
In particular, $M_n(E)=M_n\otimes_{\min} E$ for every operator space
$E$. The tensor norm $\min$ in the category of operator spaces will
play the role of $\epsilon$ in the classical theory of tensor norms.
In particular it is injective, in the sense that if $E\subset X$ and
$F\subset Y$, then $E\otimes_{\min} F\subset X\otimes_{\min} Y$ as
operator spaces. The analogue of the $\pi$ tensor norm is the
projective tensor norm, defined as
$$\|u\|_{M_n(E\otimes^{\wedge} F)}=\inf\{\|\alpha\|_{M_{n,lm}}\|x\|_{M_l(E)}\|y\|_{M_m(F)}\|\beta\|_{M_{lm,n}}:u=\alpha(x\otimes y)\beta\},$$
where $u=\alpha(x\otimes y)\beta$ means the matrix product
$$u=\sum_{rsijpq}\alpha_{r,ip} \beta_{jq,s} |r\>\<s| \otimes x_{ij}\otimes y_{pq}  \in M_n\otimes E\otimes F.$$

Both tensor norms ${\wedge}$ and $\min$ are associative and
commutative and they share the duality relations of their classical
counterparts $\pi$ and $\epsilon$. In fact, for finite dimensional
operator spaces we have the natural completely isometric
identifications
\begin{equation}\label{eq:dual-op-space}
(E\otimes^{\wedge} F)^*=CB^2(E,F; \C)=CB(E,F^*)=E^*\otimes_{\min}
F^*,
\end{equation}
where, given an operator space $E$, we define its dual operator
space $E^*$ via the identification $M_n(E^*)=CB(E,M_n)$.

\

Depending on the way one embeds a Banach space inside $\B(H)$, the
same Banach space can have a completely different operator space
structure. This happens even in the simplest example: the case of a
Hilbert space. A trivial way of embedding a finite dimensional
Hilbert space $\ell_2^n$ inside some $\B(H)$ is to put it into the
first column (resp. row) of $M_n$, that is, $|i\>\mapsto |i\>\<0|$
(resp. $|i\>\mapsto |0\>\<i|$). This gives us the column operator
space $C_n$ (resp. the row operator space $R_n$). It is trivial to
verify $$\left\|\sum_i A_i\otimes |i\> \right\|_{M_m\otimes_{\min}
R_n}=\left\|\sum_i A_iA_i^{\dagger}\right\|^{\frac{1}{2}}, \quad
\quad \left\|\sum_i A_i\otimes |i\> \right\|_{M_m\otimes_{\min}
C_n}=\left\|\sum_i A_i^{\dagger}A_i\right\|^{\frac{1}{2}}.$$

We can also define the intersection of these two operator spaces
$\rc_n=R_n\cap C_n$, where given two operator spaces $E,F$
\cite[page 55]{Pisierbook}
$$\|\cdot\|_{M_n\otimes_{\min} E\cap F}= \max\{\|\cdot\|_{M_n\otimes_{\min} E}, \|\cdot \|_{M_n\otimes_{\min} F}\}.$$
We will denote by $\rc^2_n$ to $\rc_n\otimes_{\min}\rc_n$. We have
the following concrete expressions
$$\left\|\sum_i A_i\otimes |i\> \right\|_{M_m\otimes_{\min}
\rc_n}=\max\left\{\left\|\sum_i
A_iA_i^{\dagger}\right\|^{\frac{1}{2}},\left\|\sum_iA_i^{\dagger}A_i\right\|^{\frac{1}{2}}\right\},$$
\begin{align}\label{eq:rc2}
\left\|\sum_{ij} A_{ij}\otimes |ij\> \right\|_{M_m\otimes_{\min}
\rc_n^2}=&\max\left\{\left\|\sum_i
A_iA_i^{\dagger}\right\|^{\frac{1}{2}},\left\|\sum_iA_i^{\dagger}A_i\right\|^{\frac{1}{2}},\right.\\
\nonumber & \left. \left\|\sum_{ij} A_{ij}\otimes
|i\>\<j|\right\|_{M_m\otimes_{\min}M_n},\left\|\sum_{ij}
A_{ij}\otimes |j\>\<i|\right\|_{M_m\otimes_{\min}M_n} \right\}.
\end{align}

The first estimate is trivial and the second one can be easily
derived by applying the following isometric identifications
\cite[page 163]{EffrosRuan}
\begin{align*}
R_n\otimes_{\min} R_n= R_{n^2}, \quad \quad C_n\otimes_{\min}
C_n=C_{n^2},\quad\quad C_n\otimes_{\min} R_n=M_n
\end{align*}
and decomposing $(R_n\cap C_n)\otimes_{\min} (R_n\cap C_n)=
(R_n\otimes_{\min} R_n) \cap (R_n\otimes_{\min} C_n)\cap
(C_n\otimes_{\min} R_n)\cap (C_n\otimes_{\min} C_n)$ \cite[page
55]{Pisierbook}.

With
(\ref{eq:rc2}) it is trivial to verify that
\begin{lem}\label{lem:estimates-rc}
$$\left\|\sum_{ij=1}^N |i\>\<j|\otimes  |ij\> \right\|_{M_N(\rc_{N^2})}=\sqrt{N}.$$
\end{lem}

Moreover, we have the canonical completely isometric identifications
$$R_n^*=C_n,\quad C_n^*=R_n,$$
and the formal identities $R_n\lra \rc_n^*$, $C_n\lra \rc_n^*$ are
completely contractive.

\

The connection with Theorem \ref{thm:marius} will be made by the
following non-commutative Khintchine's inequality, proved by
Lust-Picard and Pisier in \cite{Lust-Picard} (see also
\cite{Pisierbook}, Sec. 9.8).

\

Before stating it, we need to give an alternative view of the
Rademacher functions. Given the group of signs $D_n=\{-1,1\}^n$ and
the normalized Haar measure on it $\mu_n$, we define the $i$-th
Rademacher function $\epsilon_i:D_n\lra \R$ as the $i$-th coordinate
function. If we call $E_n=\spanned\{\epsilon_i:1\le i\le n\}\subset
L_1(D_n,\mu_n)=\ell_1^{2^n}$ (where $\ell_1^m=(\C^m,\|\cdot\|_1)$)
then
\begin{thm}[Lust-Picard/Pisier]\label{thm:lust-picard}
The canonical identity map
$id:\rc_n^*\lra E_n$ given by $|i\>\mapsto \epsilon_i$ verifies that
$\|id\|_{cb}\|id^{-1}\|_{cb}\le C$, where $C$ is some universal
constant, and the operator space structure on $\ell_1^{2^n}$ is
determined by $\ell_1^{2^n}=(\ell_{\infty}^{2^n})^*$.
\end{thm}

\

Among all possible operator space structures for a finite
dimensional Hilbert space $\ell_2^m$, there is one that is the
{\it minimal} in the sense that every bounded operator with range
$\min(\ell_2^m)$ is always completely bounded. This is exactly the
operator space structure inherited from the embedding
$\ell_2^m\lra \ell_{\infty}(S^{m-1})$ given by $|i\>\mapsto f_i$
where $f_i(|\phi\>)=\<\phi|i\>$ for every $|\phi\>$ in the unit
sphere $S^{m-1}$. There are some properties we will need about
$\min(\ell_2^m)$. The first one is that $\min(\ell_2^m)$ is a
$1$-exact operator space in the following sense
(\cite{Pisierbook}, Chap. 17):

\

An operator space $E$ is called $\lambda$-exact if, given any
$C^*$-algebra $A$ and any (closed two-sided) ideal $\I\subset A$,
the complete contractive map $Q:\frac{A\otimes_{\min}
E}{\I\otimes_{\min}E}\lra \frac{A}{\I}\otimes_{\min}E$ verifies that
$\|Q^{-1}\|\le\lambda$. In particular, for $\min(\ell_2^m)$, $Q$ is
a complete isometry.

\

Moreover, for any operator space $E$,
$E\otimes_{\min}\min(\ell_2^m)=E\otimes_{\epsilon}\ell_2^m$ as
Banach spaces. With this and (\ref{eq:rc2}) one can finally obtain
\begin{lem}\label{lem:estimates-min}
\begin{align*}
\|\sum_{ij} |ij\>\otimes
|ij\>\|_{\rc^2_n\otimes_{\min}\min(\ell_2^{n^2})}\le 1
\end{align*}
\end{lem}

\subsection*{Random matrices and Wassermann's construction}

We start with the following application of Chevet's inequality.
Many of the ideas behind the proof come from the seminal work
\cite[Chapter V]{MarcusPisier}. We essentially follow here
\cite{MariusHabi} which is only available on a preprint server.
Therefore we include a complete proof of the statement for
convenience of the reader.

\begin{lem}\label{rand1} Let $n, N\in {\mathbb N}$  and $\mathbb{U}_N^n$ the $n$-fold product of
the unitary group equipped with the normalized Haar measure. Then
\[ \ez \sup_{\sum|\lambda_i|^2\le 1} \|\sum_{i=1}^n \lambda_i
U_i\|_{M_N}\le 32\pi (1+\sqrt{\frac{n}{4N}}) \,. \]
\end{lem}

\begin{proof} We recall Chevet's inequality. For Banach spaces
$E,F$, \cite[Theorem 43.1]{Tomczak}
\begin{align}\label{chevet}
 \ez \|\sum_{s,t} g_{s,t} x_s\otimes  y_t\|_{E\otimes_{\varepsilon}F}
 &\le b\|(x_s)_s\|_2^{\w}\ez\|\sum_{t} g_ty_t\|_F+b\|(y_t)_t\|_2^{\w} \ez \|\sum_s g_s
 x_s\|_E \;.
 \end{align}
Here $g_{s,t}$ are independent normalized real gaussian random
variables, $b=1$ if the spaces are real whereas $b=4$ if they are
complex, and we recall that
$$\|(x_s)_s\|_2^{\w}=\sup \{\left(\sum_s |x^*(x_s)|^2\right)^{\frac{1}{2}}\; |\; x^*\in E^*, \|x^*\|\le 1\}   \;.$$
Let us apply this twice to get
\begin{align*}
& \ez \|\sum_{i=1,...,n,k,l=1,...,N} g_{ikl} \,  |i\>\otimes
|k\>\otimes
 |l\>\|_{\ell_2^n\otimes_{\varepsilon} \ell_2^N\otimes_{\varepsilon} \ell_2^N}
 \\
 &\, \le 4\|(|i\>)_i\|_2^{\w}\,   \ez\|\sum_{kl} g_{kl} |k\>\otimes |l\>\|_{\ell_2^N\otimes_{\varepsilon}
 \ell_2^N} + 4\ez\|\sum_{i=1}^n g_i |i\>\|_2 \, \|(|k\>\otimes
 |l\>)_{kl}\|_2^{\w}\\
 &\, \le
 8 \ez\|\sum_{k=1}^N g_k |k\>\|_2 + 4\sqrt{n}\le
 8\sqrt{N}+4\sqrt{n} \,  .
 \end{align*}

In order to transform this to unitaries we replace $g_{ijk}$ by
complex gaussians $\tilde{g}_{ijk}=\frac{g_{ijk}+ {\bf i}
g'_{ijk}}{\sqrt{2}}$. This gives an additional factor $\sqrt{2}$.
Then, following \cite[Lemma 3.2.1.5]{MariusHabi},we obtain (see
below for the details)
\begin{align}\label{eq:lema-chevet}
&\frac{1}{N} \ez \| \sum_{jk}
\tilde{g}_{jk} \,  |j\rangle \langle k|
\|_{S_1^N}
 \ez \|\sum_{i=1}^n |i\>\otimes U_i\|_{\ell_2^n\otimes_\epsilon
 M_N}\\
\nonumber&  \, \le \ez \|\sum_{i=1,...,n;\, j,k=1,...,N}
\tilde{g}_{ijk} |i\>\otimes |j\>\otimes
|k\>\|_{\ell_2^n\otimes_{\varepsilon}
\ell_2^N\otimes_{\varepsilon}
 \ell_2^N}
 \, \le\,  8\sqrt{2 N} +4\sqrt{2n}.
 \end{align}
Finally, we have $ \ez \| \sum_{jk} \tilde{g}_{jk} \,  |j\rangle
\langle k| \|_{S_1^N}\ge \frac{N^{3/2}}{\pi\sqrt{2}}$. To see this
it is enough to show that for {\it real} gaussians and {\it real}
Hilbert spaces $\ell_2^N$ we get
$$\ez \| \sum_{jk}
g_{jk} \,  |j\rangle \langle k|
\|_{\ell_2^N\otimes_\pi\ell_2^N}\ge
\frac{N^{3/2}}{\pi}.$$ Recall from above the notation
$D_n=\{-1,1\}^n$,  $\mu_n$ the Haar measure on $D_n$ and
$\epsilon_i$ the $i$-th Rademacher function. It is a simple
exercise \cite[Section 8.7]{Defant} to verify that
\begin{align*}
\ez
[(\sum_{jk}|g_{jk}|^2)^\frac{1}{2}]&=\int_\Omega\|\sum_{jk}g_{jk}(\omega)|j\>\<k|\|_{\ell_2^N\otimes_{\Delta_2}\ell_2^N}
d\P(\omega)\\
 & \ge\sqrt{\frac{2}{\pi}}
 \int_{D_{N^2}}\|\sum_{jk}\epsilon_{jk}(s)\;|j\>\<k|\|_{\ell_2^N\otimes_{\Delta_2}\ell_2^N}d\mu_{N^2}(s)=\sqrt{\frac{2}{\pi}}N.
\end{align*}
Using the duality $(\ell^N_2\otimes_\varepsilon
\ell^N_2)^*=\ell^N_2\otimes_\pi \ell^N_2$ and H\"older's
inequality this implies
 \begin{align*}
 \sqrt{\frac{2}{\pi}}N &\leq \ez
 [(\sum_{jk}|g_{jk}|^2)^\frac{1}{2}]=\ez [|\langle
 \sum_{jk}g_{jk}|j\>\<k|,\sum_{st}{g}_{st}|s\>\<t|\rangle|
 ^\frac{1}{2}]\\
 & \leq (\ez
 \|\sum_{jk}g_{jk}|j\>\<k|\|_{\ell^N_2\otimes_\varepsilon
 \ell^N_2})^\frac{1}{2}(\ez
 \|\sum_{jk}g_{jk}|j\>\<k|\|_{\ell^N_2\otimes_\pi
 \ell^N_2})^\frac{1}{2}.
 \end{align*}
Now, using Chevet's inequality again, we know that $(\ez
\|\sum_{jk}g_{jk}|j\>\<k|\|_{\ell^N_2\otimes_\varepsilon
\ell^N_2})^\frac{1}{2}\leq \sqrt{2}N^\frac{1}{4}$. Thus we have
$$\ez \|\sum_{jk}g_{jk}|j\>\<k|\|_{\ell^N_2\otimes_\pi
\ell^N_2}\geq \frac{1}{\pi}N^\frac{3}{2}.$$

\

So it only remains to the first inequality  in
(\ref{eq:lema-chevet}). We include the argument given in
\cite[Lemma 3.2.1.5]{MariusHabi} for completeness. Let
$(U_i)_{i=1}^n\subset \mathbb{U}_N$ be a sequence of unitary
matrices. We consider left multiplication $L$ with respect to
block diagonal of $U_i's$, namely $L:\mathbb{C}^{nN^2}\lra
\mathbb{C}^{nN^2}$ defined by
$$(x_{jk}^i)_{jk} \longmapsto (U_i \circ (x_{jk}^i)_{jk}), \; \; \forall i$$
as well as the corresponding right multiplication. They are unitary
operations on $\mathbb{C}^{nN^2}$ and therefore leave the complex
gaussian density invariant.

\

Given a sequence of random matrices $G^i(\omega)$ with independent
normalized complex gaussian entries, that is
$G^i_{jk}(\omega)=\tilde{g}_{jk}^i(\omega)$, we denote by
$\tau^i(\omega)$ the sequence of singular values of $G^i(\omega)$,
in the sense that there are unitaries $U^i(\omega)$, $V^i(\omega)$
with
$$G^i(\omega)=U^i(\omega)D_{\tau^i(\omega)}V^i(\omega).$$

\

We denote by $\Pi$ the Haar measure on the group $G$ of sequences of
permutations $G=(Perm\{1,\cdots ,N\})^n$ and $M_{\pi}$ the
permutation matrix $M_{\pi}(|i\>)=|\pi(i)\>$. For $i=1,\cdots ,n$
and a diagonal operator $D_{\tau^i}$, we have
$$\int_GM_{\pi^i}D_{\tau^i}
M_{(\pi^i)^{-1}}d\Pi=(\frac{1}{N}\sum_l\tau_l^i)
id_{\mathbb{C}^N}.$$ Now, let $C\subset \mathbb{C}^{nN^2}$ be a
finite set. Calling $\mu$ to the Haar measure in $\mathbb{U}_N^n$ we
get
$$\int_\Omega\sup_{(x_{ijk})_{ijk}\in
C}|\sum_{ijk}x_{ijk}\tilde{g}_{jk}^i(\omega)|d\mathbb{P}(\omega)=\int_{\mathbb{U}^n_N\times
\mathbb{U}^n_N}\int_\Omega\sup_{(x_{ijk})_{ijk}\in
C}|\sum_{ijk}x_{ijk}(U^iG^i(\omega)V^i)_{jk}|d\mathbb{P}(\omega)d\mu
d\mu$$
$$=\int_\Omega\int_{\mathbb{U}^n_N\times
\mathbb{U}^n_N}\sup_{(x_{ijk})_{ijk}\in
C}|\sum_{ijk}x_{ijk}(U^iU^i(\omega)D_{\tau^i(\omega)}V^i(\omega)V^i)_{jk}|d\mu
d\mu d \mathbb{P}(\omega)$$
By the invariance of the Haar measure
we can write further
 \begin{align*}
& =\int_\Omega\int_G\int_{\mathbb{U}^n_N\times
\mathbb{U}^n_N}\sup_{(x_{ijk})_{ijk}\in C}|\sum_{ijk}x_{ijk}(U^i
M_{\pi^i} D_{\tau^i(\omega)} M_{(\pi^i)^{-1}} V^i)_{jk}|d\mu d\mu
d \Pi d \mathbb{P}(\omega)\\
 &\geq \int_{\mathbb{U}^n_N\times
 \mathbb{U}^n_N}\sup_{(x_{ijk})_{ijk}\in
C}|\sum_{ijk}x_{ijk}(U^i(\int_\Omega\int_G M_{\pi^i}
D_{\tau^i(\omega)} M_{(\pi^i)^{-1}} d\Pi
d\mathbb{P}(\omega))V^k)_{jk}|d\mu d\mu\\
&=\int_\Omega
 \frac{1}{N}\sum_l\tau_l^1(\omega)d\mathbb{P}(\omega)\int_{\mathbb{U}^n_N\times
 \mathbb{U}^n_N}\sup_{(x_{ijk})_{ijk}\in
 C}|\sum_{ijk}x_{ijk}(U^iV^i)_{jk}|d\mu d\mu\\
& =\int_\Omega
\frac{1}{N}\sum_l\tau_l^1(\omega)d\mathbb{P}(\omega)\int_{\mathbb{U}^n_N}\sup_{(x_{ijk})_{ijk}\in
C}|\sum_{ijk}x_{ijk}u_{jk}^i|d\mu.
\end{align*}
Now, since
$\int_\Omega\frac{1}{N}\sum_j\tau_j^1(\omega)d\mathbb{P}(\omega)=\frac{1}{N}\ez
\| \sum_{jk} \tilde{g}_{jk}\,  |j\rangle \langle k| \|_{S_1^N}$,
and taking $C$ approaching the unit ball of
$(\ell_2^n\otimes_\epsilon \ell_2^N \otimes_\epsilon \ell_2^N)^*=
(\ell_2^n\otimes_\epsilon M_N)^*$, we get (\ref{eq:lema-chevet}).

\end{proof}

\

We will use a theorem of Voiculescu in order to obtain a state of
the form of Theorem \ref{thm:marius} (defined by random unitary
matrices). We will need to define some previous concepts.

For a countable discrete group we recall that the left regular
representation $\lambda:G\to B(\ell_2(G))$ is given by
$\la(g)\delta_h=\delta_{gh}$. Here $(\delta_h)$ stands for the unit
vector basis in $\ell_2(G)$. Then $C_{red}(G)$, the norm closure of
the linear span of $\la(G)$, is called the reduced $C^*$-algebra of
$G$. The reduced $C^*$-algebra sits in the von Neumann algebra
$VN(G)=\la(G)''$. The normal trace $\tau$ on $VN(G)$ is given by
$\tau(x)=(\delta_e,x\delta_e)$.

\

For the free group $F_n$ in $n$ generators $g_1,\ldots, g_n$, the
reduced $C^*$-algebra can be realized by random unitaries in the
following sense: Let $(U_{N,i})_{i=1}^n$ be random unitaries in
$\prod_N \mathbb{U}_N^n$, endowed with the Haar measure and
$\tau_N$ the normalized trace on $\mathbb{U}_N$
($\tau_N(x)=\frac{1}{N}tr_N(x)$) for each $N$. According to
\cite[Theorem 4.3.3]{Voiculescu}, we have that
\begin{equation}\label{voi}
  \lim_N \tau_N(U_{N,i_1}^{\varepsilon_1}\cdots U_{N,i_m}^{\varepsilon_m})
 = \tau(\lambda(g_{i_1})^{\varepsilon_1} \cdots
 \lambda(g_{i_m})^{\varepsilon_m})) \end{equation}
holds almost everywhere, for every string $(i_1,...,i_m)\in
\{1,...,n\}^m$ and $\varepsilon_j=\pm 1$. Here $\tau$ is the
normalized trace on the von Neumann algebra $\lambda(F_n)''$. This
means that the right hand expression is $1$ if and only if
$g_{i_1}^{\varepsilon_1}\cdots g_{i_m}^{\varepsilon_m}$ is the
trivial word $e$ (after cancelation). In all the other cases, we
obtain $0$. We will use this result in a more quantitative way as
follows:

\ We define the set $\Omega:= \{\omega=\{a_1,\cdots ,a_k\}| k\in
\mathbb{N}, 1 \not \in \{a_1,\cdots ,a_k\}\subset F_n\}$. Given $1
\not =a\in F_n$, Voiculescu's theorem \cite[Theorem
4.3.3]{Voiculescu} tells us that
$$\lim_N \mu_N(\{(U_1,\cdots ,U_n)| \tau_N(\pi_{U_1,\cdots
,U_n}(a))<\frac{1}{k}\})=1,$$where we call $\pi_{U_1,\cdots
,U_n}:F_n\lra M_N$ to the representation of $F_n$ uniquely
determined by $g_i\lra U_i$, $i=1,\cdots ,n$. Given
$\omega=\{a_1,\cdots ,a_k\}\in \Omega$ of cardinality $k$, we deduce
the existence of $N_\omega$ such that
\begin{equation}\label{expext}
\mu_{N_\omega}(\{(U_1,\cdots ,U_n)|
\tau_{N_{\omega}}(\pi_{U_1,\cdots ,U_n}(a_i))<\frac{1}{k}\; \;
\forall i=1,\cdots ,k\})>\frac{1}{2}.
\end{equation}

Now, we know by Lemma \ref{rand1} that
\begin{equation}\label{estim}\mathbb{E}\sup_{\sum|\lambda_i|^2\le 1} \|\sum_{i=1}^n \lambda_i
U_i\|_{M_{N_\omega}}\le 32\pi
(1+\sqrt{\frac{n}{4N_\omega}}).\end{equation}

As a consequence of Chebychev's inequality, for every $\omega\in
\Omega$ there exists a sequence $(U_{N_\omega,j})_{j=1}^n\in
\mathbb{U}_{N_\omega}^n$ which satisfies both (\ref{expext}) and
(\ref{estim}) (multiplying by $2$ the bound of (\ref{estim})). These
sequences of random unitary matrices will be crucial in our
construction and will be fixed from now on.

We simplify the notation a bit more: For every $\omega\in \Omega$ we
call $\pi_\omega$ to $\pi_{U_{N_\omega, 1},\cdots ,U_{N_\omega,n}}$,
and $\tau_\omega: M_{N_\omega}\lra \mathbb{C}$ to the normalized
trace $\tau_\omega(x)=\tau_{N_\omega}(x)$.

We follow now a construction of Wassermann \cite{Wa} to obtain a
representation of the reduced $C^*$-algebra $C_{red}(F_n)$. We fix
an ultrafilter $\U$ on $\Omega$ refining the sets
$$\Omega_{\omega=\{a_1,\cdots ,a_k\}}:= \{\{b_1,\cdots ,b_n\}\subset F_n| k\leq n, \omega \subseteq \{b_1,\cdots ,b_n\}\}.$$

Then, we have that for $a\neq 1$,
\begin{equation}\label{trace}
\lim_\U\tau_\omega(\pi_\omega(a))=0.
\end{equation}

We consider the space $\ell_\infty(\Omega, M_{N_\omega})$, and we
define the (closed two-side) ideal
$$\I=\{(x_\omega)_\omega\in \ell_\infty(\Omega, M_{N_\omega}):\lim_\U \tau_\omega(x_\omega^\dagger
x_\omega)=0\}.$$ We also consider the quotient
$M_{\U}=\ell_\infty(\Omega, M_{N_\omega})/\I$, which is a finite
von Neumann algebra.

Finally, we consider the group representation $\pi:F_n\to M_\U$,
defined by $$\pi(a)=(\pi_\omega(a))_\omega+\I_\U.$$

\begin{obs}
It is trivial to check that we can do the same construction taking
$\bar{\pi}_\omega:F_n\to M_{N_\omega}$, defined by
$\bar{\pi}_\omega(g_i)=\overline{U}_{N_\omega,i}$.

\end{obs}

This construction was done in \cite[Sec. 1]{Wa}. Following this
work, and using the crucial property (\ref{trace}), the next
theorem follows directly

\begin{thm}[Wassermann] $\pi$ extends to an injective $^*$-homomorphism on $\lambda(F_n)''$,
which we also call $\pi$.
\end{thm}

Following the same argument, we obtain a result for the product
$F_n^2=F_n\times F_n$ of the free group. Here we use
$\ell_{\infty}(\Omega, M_{N_\omega}\otimes M_{N_\omega})$ and the
ideal $\I^2_\U=\{(x_\omega)_{\omega}: \lim_{\U}
\tau_{N_\omega^2}(x_\omega^\dagger x_\omega)=0\}$ and we write
$M^2_\U=\ell_{\infty}(\Omega, M_{N_\omega}\otimes
M_{N_\omega})/\I^2_\U$ and $\tau^2_\omega$ for the corresponding
trace $\tau^2_\omega(x_\omega)=\tau_{N_\omega^2}(x_\omega^\dagger
x_\omega)$.

As before, we define $\pi^2:F_n\times F_n\lra M^2_{\U}$ by
$$\pi^2(a_1,a_2) = (\bar{\pi}_\omega(a_1)\ten
\pi_\omega(a_2))_\omega+\I^2_\U,$$ and again,  using that \[
\lim_{\U}\tau_\omega^2((\bar{\pi}_\omega(a_1)\ten \pi_\omega(a_2)))
 = \lim_{\U}\tau_\omega(\bar{\pi}_\omega(a_1))\tau_\omega(\pi_\omega(a_2)) =
\delta_{a_1,1}\delta_{a_2,1}\, ,  \] we can get the analogue of
Wassermann's result:

\begin{thm}\label{Was} $\pi^2$ extends to an injective $^*$-homomorphism between
$\lambda(F_n\times F_n)''$ and $M^2_\U,$ which we also call $\pi^2$.
\end{thm}

The next proposition will be crucial in the proof of the main
theorem.

\begin{prop}\label{lem:def-SN}
There exist matrices $T^{N_\omega}_{ii'}\in M_{N_\omega^2}$ such
that, if we define
$S^{N_\omega}_{ii'}=\overline{U}_{N_\omega,i}\otimes U_{N_\omega,i'}
+ T_{ii'}^{N_\omega}$, we have
\begin{align*}
&\sup
\{|\sum_{ii'jj'kk'}a_{ii'}b_{jj'}c_{kk'}\<kk'|S^{N_\omega}_{ii'}
|jj'\>|\; ; \sum_{ii'}|a_{ii'}|^2\le 1, \sum_{jj'}|b_{jj'}|^2\le
1,
\sum_{kk'}|c_{kk'}|^2\le 1\}\le 5 \quad \text{ and } \\
&\lim_\U \tau_{\omega}^2(U_{N_\omega,i}^T\otimes
U_{N_\omega,i'}^\dagger S_{hh'}^{N_\omega})=\delta_{ih}\delta_{i'h'}
\; .
\end{align*}

\end{prop}

\begin{proof}
We call $id:\rc_n\lra C_{red}(F_n)$ (resp. $id^2:\rc_n^2\lra
C_{red}(F_n\times F_n)$) to  $id(|i\>)=g_i$ (resp.
$id^2(|ij\>)=(g_i,g_j)$). By \cite[Theorem 9.7.1]{Pisierbook}
$\|id\|_{cb}\le 2$ and then $\|id^2\|_{cb}\le 4$ (just by tensoring
with $\otimes_{\min}$, since $C_{red}(F_n)\otimes_{\min}
C_{red}(F_n)\subset C_{red}(F_n\times F_n)$ \cite[Chapter
8]{Pisierbook}).

\

We consider the map $\pi^2 id^2:\rc^2_n\to M^2_\U$ and the
amplification
\begin{align*}
\pi^2 id^2\otimes \1_{n^2}&: \rc^2_n\otimes_{\min}
\min(\ell_2^{n^2})\lra \frac{\ell_{\infty}(M_{N_\omega}\ten
M_{N_\omega})}{\I_\U^2}\otimes_{\min} \min(\ell_2^{n^2}).
\end{align*}

Using that any $*$-homomorphism (in particular $\pi^2$) is
completely contractive, that $\min(\ell_2^m)$ is a $1$-exact
operator space and Lemma \ref{lem:estimates-min}, there exists a
lifting
\begin{align*}
Z_{N_\omega}&= \sum_{ii'} \left(\overline{U}_{N_\omega,i}\otimes
U_{N_\omega,i'}+T_{ii'}^{N_\omega}\right)\otimes |ii'\> \in
M_{N_\omega^2}\otimes_{\min}
\min(\ell_2^{n^2})
\end{align*}
with $T_{ii'}^{N_\omega} \in \I^2_{\U}$ and
$\sup_\omega\|Z_{N_\omega}\|\le 5$. Now we use that
$M_{N_\omega^2}\otimes_{\min} \min(\ell_2^{n^2})=
M_{N_\omega^2}\otimes_{\epsilon} \ell_2^{n^2}$ to show that
$$\|Z_{N_\omega}\|=\sup
\{|\sum_{ii'jj'kk'}a_{ii'}b_{jj'}c_{kk'}\<kk'|S^N_{ii'} |jj'\>|\;
; \sum_{ii'}|a_{ii'}|^2\le 1, \sum_{jj'}|b_{jj'}|^2\le 1,
\sum_{kk'}|c_{kk'}|^2\le 1\}\le 5.$$

\

To conclude it is enough to show that
\begin{equation}\label{eq:limit-ultra}
\lim_{\U} \tau_\omega^2((U_{N_\omega,i}^T\ten
U_{N_\omega,i'}^\dagger)(\overline{U}_{N_\omega,h}\otimes
U_{N_\omega,h'}+T_{hh'}^{N_\omega})) =
\delta_{i,h}\delta_{i',h'}.
\end{equation}
Indeed, by (\ref{trace}) we have
\[ \lim_{\U} \tau_\omega^2((U_{N_\omega,i}^T\ten U_{N_\omega,i'}^\dagger)(\overline{U}_{N_\omega,h}\otimes
U_{N_\omega,h'}))=
\delta_{i,h}\delta_{i',h'}. \]
Moreover, since $(T_{hh'}^{N_\omega})\in \I_\U^2$ we deduce
\begin{align*}
\lim_{\U} |\tau_\omega^2((U_{N_\omega,i}^T\ten U_{N_\omega,i'}^\dagger)T_{hh'}^{N_\omega})| &\le \lim_{\U} \tau_\omega^2((U_{N_{\omega},i}^T\ten
 U_{N_\omega,i'}^\dagger)(\overline{U}_{N_\omega,i}\ten U_{N_\omega,i'}))^{1/2} \tau_\omega^2((T_{hh'}^{N_\omega})^\dagger(T_{hh'}^{N_\omega}))^{1/2}
\\
&=\lim_{\U} \tau_\omega^2((T_{hh'}^{N_\omega})^\dagger(T_{hh'}^{N_\omega}))^{1/2} =0
\,.
\end{align*}

\end{proof}

\

\begin{obs} The operators $T_{ii'}^N$ are highly non-trivial. This can
be seen by noticing that $\|\sum_{i=1}^n \overline{U}_{N,i}\otimes
U_{N,i}\|_{M_{N^2}}= n$. This is by factor $\sqrt{n}$ larger than
$\|\sum_{i=1}^n \overline{U}_{N,i}\otimes U_{N,i}+T_{ii}^N\|
\le 5 \sqrt{n}$,
guaranteed from the Wassermann lifting.
\end{obs}

\subsection*{Proof of the result}

We define the (unnormalized) state $|\psi_{N_\omega}\>=
\frac{1}{\sqrt{nN_{\omega}}} \sum_{1\le i\le n\; 1\le j,k\le
N_\omega} \<j|U_{N_\omega,i}^\dagger|k\> |ijk\>$. We know that these
matrices verify the estimate from Lemma \ref{rand1} and hence
$$\||\psi_{N_\omega}\>\|_{\ell_2^n\otimes_\epsilon M_{N_\omega}}\preceq \frac{1}{\sqrt{nN_\omega}},$$
which means that $\<\psi_{N_\omega}|\psi_{N_\omega}\>\preceq 1$.

\

We define the trilinear form $v_{N_\omega}:\ell_2^{n^2}\times
\ell_2^{N_\omega^2}\times \ell_2^{N_\omega^2}\lra \C$ by
$$v_{N_\omega}(|ii'\>,|jj'\>,|kk'\>)=\<kk'|S_{ii'}|jj'\>.$$
Thanks to Proposition \ref{lem:def-SN}, $\|v_{N_\omega}\|\le 5$. If
we call $q=id^*$ where $id$ is the map given in Theorem
\ref{thm:lust-picard}, we define $T$ via the diagram
$$
\begin{diagram} \dgARROWLENGTH=3em
\node{\ell_\infty^{2^{n^2}}\times \ell_\infty^{2^{N_\omega^2}}\times
\ell_\infty^{2^{N_\omega^2}}}\arrow{s,l}{q\otimes q\otimes
q}\arrow{se,t}{T}\\
\node{\ell_2^{n^2}\times \ell_2^{N_\omega^2}\times
\ell_2^{N_\omega^2}}\arrow{e,t}{v_{N_\omega}} \node{\C}
\end{diagram}
$$
It is clear that $\|T\|\preceq 1$. Moreover, since
$q:\ell_{\infty}^M\lra \rc_m$ is a complete quotient (Theorem
\ref{thm:lust-picard}), there exist $b\in
M_n(\ell_\infty^{2^{n^2}})$ and $\hat{b}\in
M_{N_\omega}(\ell_\infty^{2^{N_\omega^2}})$ such that
\begin{align*}
&(\1_n\otimes q)(b)=\frac{1}{\sqrt{n}}\sum_{ii'=1}^n |i\>\<i'| \otimes |ii'\> \\
&(\1_{N_\omega}\otimes
q)(\hat{b})=\frac{1}{\sqrt{N_\omega}}\sum_{jj'=1}^{N_\omega}
|j\>\<j'| \otimes |jj'\>\\
&\|b\|, \|\hat{b}\| \preceq 1 \quad \text{ (by Lemma
\ref{lem:estimates-rc})}
\end{align*}

It remains to be proven that (for some $N_\omega$) $$
\left|\<\psi_{N_\omega}|T_{n,N_\omega,N_\omega}(b,\hat{b},\hat{b})|\psi_{N_\omega}\>\right|\succeq
\sqrt{n}.$$

To see this we notice
\begin{align*}
&\left|\<\psi_{N_\omega}|T_{n,N_\omega,N_\omega}(b,\hat{b},\hat{b})|\psi_{N_\omega}\>\right|=\frac{1}{N_\omega\sqrt{n}}
\left|\<\psi_{N_\omega} |\left(\sum_{ii'jj'kk'}
v_{N_\omega}(|ii'\>,|jj'\>,|kk'\>)\;
|i\>\<i'|\otimes |j\>\<j'|\otimes |k\>\<k'| \right)|\psi_{N_\omega}\>\right|=\\
&\frac{1}{N_\omega^2n\sqrt{n}} \left|\sum_{ii'jj'kk'} \<kk'
|S_{ii'}^{N_\omega}|jj' \> \overline{\<j|U_{N_\omega,i}^\dagger|k\>}
\<j'|U_{N_\omega,i'}^\dagger|k'\>\right|=\frac{1}{N_\omega^2n\sqrt{n}}
\left|\sum_{ii'}\tr((U_{N_\omega,i}^T\otimes
U_{N_\omega,i'}^\dagger)S^{N_\omega}_{ii'})\right|=
\\
&\frac{1}{n\sqrt{n}} \left|\tr\left( \sum_{ii'hh'}
\tau_{\omega}^2(U_{N_\omega,i}^T\otimes U_{N_\omega,i'}^\dagger \;
S_{hh'}^{N_\omega}) |ii'\>\<hh'|\right)\right|\underset{\U}{\lra}
\sqrt{n},
\end{align*}
since, by Proposition \ref{lem:def-SN}, $$\lim_{\U} \sum_{ii'hh'}
\tau_{N_\omega^2}(U_{N_\omega,i}^T\otimes U_{N_\omega,i'}^\dagger \;
S_{hh'}^{N_\omega}) |ii'\>\<hh'|= id_{\ell_2^{n^2}}.$$

The result follows trivially.

\

The optimality part is a trivial consequence of the following
\begin{prop}
For any $N$ and any linear map $v:\ell_{\infty}^N\lra
\ell_1^N\otimes_{\epsilon}\ell_1^N$, if we call $v_n$ to the
amplification $v_n=\1_n\otimes v: M_n(\ell_{\infty}^N)\lra
M_n(\ell_1^N\otimes_{\min}\ell_1^N)$, then $$\|v_n\|\preceq
\sqrt{n}\|v\|.$$
\end{prop}

\begin{proof}
We recall that $E_n$ is the linear span of the first $n$ Rademacher
functions in $L_1(D_n)$. $F_n$ will be $E_n\otimes E_n\subset
L_1(D_n\times D_n)$. By the classical Khintchine's inequalities (see
for instance \cite{Defant}, Sec. 8.5), we have that
$$\frac{1}{2}\left(\sum_{ij} |\alpha_{ij}|^2\right)^{\frac{1}{2}}\simeq\left\|\sum_{ij}\alpha_{ij}\epsilon_i\epsilon'_j\right\|_{L_1(D_n\times D_n)}.$$
Hence, the norm of the identity $id:F_n\lra S_2^n$
$(\epsilon_i\epsilon'_j\mapsto |i\>\<j|)$ is $\preceq 1$ and
therefore (recall that $\|\cdot\|_{S_2^n}\le
\sqrt{n}\|\cdot\|_{M_n}$) the norm of the adjoint map $id:M_n\lra
F_n^*=\frac{L_{\infty}(D_n\times D_n)}{F_n^{\perp}}$ is $\preceq
\sqrt{n}$. Using that the formal identities $R_n\lra \rc_n^*$,
$C_n\lra \rc_n^*$ are completely contractive and Theorem
\ref{thm:lust-picard}, we get that the identity
$id:R_n\otimes^{\wedge} C_n\lra E_n\otimes E_n\subset
L_1(D_n)\otimes^{\wedge} L_1(D_n)=L_1(D_n\times D_n)$ has completey
bounded norm $\preceq 1$. Then the adjoint map $\|id:F_n^*\lra
M_n\|_{cb}\preceq 1$.

\

Let us take now $x=\sum_{ij}|i\>\<j|\otimes x_{ij}\in
M_n(\ell_{\infty}^N)=M_n\otimes_{\epsilon}\ell_{\infty}^N$ with norm
$\le 1$. There exists a function $f\in L_{\infty}(D_n\times
D_n)(\ell_{\infty}^N)$ such that $\|f\|\preceq \sqrt{n}$ and
$$x_{ij}=\int_{D_n\times
D_n}\epsilon_i\epsilon'_jf(\epsilon,\epsilon')d\mu_n(\epsilon)d\mu_n(\epsilon').$$
For that we have used that if $Q:X\lra Y$ is an isometric quotient,
then $Q\otimes \id:X\otimes_\epsilon\ell_{\infty}^N\lra
Y\otimes_\epsilon\ell_{\infty}^N$ is also an isometric quotient (see
for instance \cite{Defant}, Sec. 4.4).

If we denote $g=id\otimes v(f)\in L_{\infty}(D_n\times
D_n)(\ell_1^N\otimes_{\epsilon} \ell_1^N)$, then
$\|g\|\preceq\sqrt{n}\|v\|$ and $$v(x_{ij})=\int_{D_n\times
D_n}\epsilon_i\epsilon'_jg(\epsilon,\epsilon')d\mu_n(\epsilon)d\mu_n(\epsilon').$$

If $Q:L_{\infty}(D_n\times D_n)\lra F_n^*=\frac{L_{\infty}(D_n\times
D_n)}{F_n^{\perp}}$ is the canonical quotient map, the composition
$idQ:L_{\infty}(D_n\times D_n)\lra M_n$ (given by
$idQ(h)=\sum_{ij}\left(\int \epsilon_i\epsilon_j'h \right)
\;|i\>\<j|$) has completely bounded norm $\preceq 1$ and then
$$\|v_n(x)\|_{M_n(\ell_1^N\otimes_{\min}\ell_1^N)}= \|idQ\otimes
\1_{\ell_1^N\otimes\ell_1^N}
(g)\|_{M_n(\ell_1^N\otimes_{\min}\ell_1^N)}\le
\|idQ\|_{cb}\|g\|_{L_{\infty}(D_n \times D_n)\otimes_{\epsilon}
(\ell_1^N\otimes_{\min}\ell_1^N)}\preceq \sqrt{n}\|v\|,$$ since, by
Grothendieck's theorem, $\ell_1^N\otimes_{\min}\ell_1^N\simeq
\ell_1^N\otimes_{\epsilon}\ell_1^N$ (see Section
\ref{sec:preliminaries}).
\end{proof}

\section{Conclusion}

We have shown that some tripartite quantum states, constructed in a
random way, can lead to arbitrarily large violations of Bell
inequalities. Moreover, and contrary to what happens with other
measures of entanglement, the GHZ state does not share this extreme
behavior. Apart from the interest of the results (in particular we
answer a long standing open question of Tsirelson) and from the
applications that can be derived (see Section \ref{sec:Mainresult}),
we think that one of the main achievements in the paper is the use
of completely new mathematical tools in this context. We hope that
the techniques and connections we have established here will provide
a better understanding of Bell inequalities in the near future. In
this direction we would like to finish with some open problems.

\subsection*{A couple of open questions}

We have proven in the text that there are {\it reasonably many}
states leading to large violations of Bell inequalities, since we
have constructed them using random unitaries. However, if we focus
on the inequalities (rather than on the states) the picture is much
less clear. Apart from seeking for an explicit form (see Remark
after Theorem \ref{mainBellThm}) one could ask the following:

\

{\bf Question 1}: {\it How many} Bell inequalities give large
violation?

\

Following the relations found in this paper, one can formulate this
question in the following quantitative way:

\

{\bf Question 1'}: Are the volumes of the unit balls of
$\ell_1^n\otimes_\epsilon\ell_1^n\otimes_\epsilon\ell_1^n$ and
$\ell_1^n\otimes_{\min}\ell_1^n\otimes_{\min}\ell_1^n$ comparable?

\

Once more Chevet's inequality gives us the right estimate for the
volume of the unit ball of
$\ell_1^n\otimes_\epsilon\ell_1^n\otimes_\epsilon\ell_1^n$. So the
question can be finally stated as

\

{\bf Question 1''}: Which is the (asymptotic) volume of the unit
ball of $\ell_1^n\otimes_{\min}\ell_1^n\otimes_{\min}\ell_1^n$?

\

Unfortunately, the techniques used in this paper do not seem to help
much to tackle this problem, and probably new ideas have to come
into play.

Another interesting question arising from the paper is the
possibility of giving highly non-local states with a simpler
structure than the ones given here. For instance, it would be nice
to know if

\

{\bf Question 2:} Can one find a diagonal state
$|\psi\>=\sum_{i=1}^D \alpha_i |iii\>$ giving unbounded violation to
a Bell inequality?

\

We have proven that the GHZ (i.e. $\alpha_i=\frac{1}{\sqrt{D}}$ for
every $i$) does not, but, interestingly enough, Question 2 is
equivalent to the following completely mathematical question

\

{\bf Question 2':} Is $S_\infty$ (the space of compact operators in
a Hilbert space) a Q-algebra with the Schur product?

\

This question, that was formulated by Varopoulos in 1975
\cite{Varopoulos}, is still open, though there has been some
progress towards its solution \cite{perez-q-algebra,LeMerdy}. A nice
exposition about Q-algebras can be found in \cite[Chapter
18]{DiJaTo}. We review here the basics to connect Questions 2 and
2'.

\

A $Q$-algebra is defined as a commutative Banach algebra isomorphic
to a quotient algebra of a uniform algebra, where a uniform algebra
is simply a closed subalgebra of the algebra of continuous functions
$C(K)$ for some compact Hausdorff space $K$. For a brief exposition
of the history and importance of this kind of algebras we refer the
reader to \cite[Chapter 18, Notes and Remarks]{DiJaTo}. A very
important step in the understanding of these algebras was made by
Davie \cite{Davie}, by proving the following criterion

\begin{thm}\label{thm:Davie}
A commutative Banach algebra $X$ is a $Q$-algebra if and only if
there is a universal constant $K$ such that
$$\|\sum_{i_1,\ldots,i_N}t_{i_1\ldots i_N} x^1_{i_1}\cdots x_{i_N}^N\|_X\le K^N \sup_{|\epsilon^j_{i_j}|=1}
\left|\sum_{i_1,\ldots,i_N}t_{i_1\ldots i_N}\epsilon_{i_1}^1 \cdots
\epsilon_{i_N}^N\right|.$$ For every choice of elements
$x^j_{i_j}\in X$ with $\|x^j_{i_j}\|\le 1$.
\end{thm}

To be precise this is not exactly the formulation made by Davie, but
one can easily obtain it following the reasonings of \cite[Prop.
18.6, Thm. 18.7]{DiJaTo}. Using Theorem \ref{thm:Davie}, we can
formalize the relation between Questions 2 and 2':
\begin{thm}
$S_{\infty}$ is a $Q$-algebra if and only if there is a universal
constant $K$ such that for any $N$ and any diagonal $N$-partite
state $|\psi\>=\sum_{i=1}^D \alpha_i |ii\cdots i\>$, the largest
violation that $|\psi\>$ can induce in a Bell inequality (with an
arbitrarily number of dichotomic observables) is bounded by $K^N$.
\end{thm}

\begin{proof}
Let us assume first that $S_{\infty}$ is a $Q$-algebra. By Theorem
\ref{thm:Davie}, for {\it real} $t_{i_1 \ldots i_N}$ and {\it
hermitian} $A^j_{i_j}\in M_D\subset S_{\infty}$ with
$\|A^j_{i_j}\|_{M_D}\le 1$ we have that
$$\|\sum_{i_1,\ldots,i_N}t_{i_1 \ldots i_N} A^1_{i_1}*\cdots *A_{i_N}^N\|_{S_\infty}\le K^N \sup_{|\epsilon^j_{i_j}|=1}
\left|\sum_{i_1,\ldots,i_N}t_{i_1\ldots i_N}\epsilon_{i_1}^1 \cdots
\epsilon_{i_N}^N\right|\le (2K)^N \sup_{\epsilon^j_{i_j}=\pm 1}
\left|\sum_{i_1,\ldots,i_N}t_{i_1 \ldots i_N}\epsilon_{i_1}^1 \cdots
\epsilon_{i_N}^N\right|,$$ where $*$ means Schur (or Hadamard)
product.

We now notice that
\begin{align}\label{eq:Schur}
\|\sum_{i_1,\ldots,i_N}t_{i_1 \ldots i_N} A^1_{i_1}*\cdots *
A_{i_N}^N\|_{S_\infty}&= \max_{|r\>} \left|\<r |\; \;
\sum_{i_1,\ldots,i_N}t_{i_1 \ldots i_N} A^1_{i_1}*\cdots *
A_{i_N}^N\;\;|r\>\right| \\ \nonumber &= \max_{\sum_i |a_i|^2=1}
\left|\sum_{i_1,\ldots,i_N}t_{i_1 \ldots i_N}\sum_{i,j}
\overline{a}_i{a_j} \<ii\cdots i|\; A^1_{i_1}\otimes \cdots \otimes A^N_{i_N}\; |jj\cdots j\>\right|\\\nonumber &=
\max_{|\psi\> \text{ diagonal}}\left|\sum_{i_1,\ldots,i_N}t_{i_1
\ldots i_N}\<\psi|\; A^1_{i_1}\otimes \cdots \otimes A^N_{i_N}\;
|\psi\>\right|.
\end{align}

For the other implication we assume by hypothesis, and using
(\ref{eq:Schur}), that
\begin{equation}\label{eq:Schur2}
\|\sum_{i_1,\ldots,i_N}t_{i_1 \ldots i_N} A^1_{i_1}*\cdots *
A_{i_N}^N\|_{S_\infty}\le K^N \sup_{\epsilon^j_{i_j}=\pm 1}
\left|\sum_{i_1,\ldots,i_N}t_{i_1 \ldots i_N}\epsilon_{i_1}^1 \cdots
\epsilon_{i_N}^N\right|\end{equation} for {\it real} $t_{i_1 \ldots
i_N}$ and {\it hermitian} $A^j_{i_j}\in M_N\subset S_{\infty}$ with
$\|A^j_{i_j}\|_{M_N}\le 1$. By splitting into real and imaginary
part it is easy to obtain (\ref{eq:Schur2}) for complex $t_{i_1
\ldots i_N}$ and arbirary matrices $A^j_{i_j}\in M_D$ of norm $1$
(maybe with a different constant $K'$). Since, given any
$\epsilon>0$, we can approximate any element $x\in S_\infty$ of
$\|x\|\le 1$ by a matrix $A\in M_D$ with $\|A\|\le 1$ and
$\|x-A\|_{S_\infty}\le \epsilon$, we obtain
$$\sup_{\|x^j_{i_j}\|\le 1}\|\sum_{i_1,\ldots,i_N}t_{i_1 \ldots i_N}
x^1_{i_1}*\cdots* x_{i_N}^N\|_{S_\infty}\le K'^N
\sup_{\epsilon^j_{i_j}=\pm 1} \left|\sum_{i_1,\ldots,i_N}t_{i_1
\ldots i_N}\epsilon_{i_1}^1 \cdots \epsilon_{i_N}^N\right|\le K'^N
\sup_{|\epsilon^j_{i_j}|= 1} \left|\sum_{i_1,\ldots,i_N}t_{i_1
\ldots i_N}\epsilon_{i_1}^1 \cdots \epsilon_{i_N}^N\right|,$$ which
finishes the proof of the theorem.
\end{proof}

\section*{Acknowledgments}

The authors are grateful to  D. Kribs and M.B. Ruskai for the
organization of the BIRS workshop {\it Operator Structures in
Quantum Information Theory}, where part of this paper was made. We
thank M. Zukowski and M. B. Ruskai for valuable comments and
acknowledge financial support from Spanish grants MTM2005-00082
and Ram\'on y Cajal.

\appendix*

\section{Proof of Theorem \ref{thm:separable}}\label{sec:appendix}

The aim of this appendix is to show again the advantages of using
the theory of tensor norms to tackle some problems on Quantum
Information. Here we provide a new proof for Theorem
\ref{thm:separable}. The key point of the proof is the following
characterization of separability \cite{Rudolph} (see also
\cite{David-sep}):

\begin{thm}
A tripartite state $\rho$ on $\C^d\otimes \C^d\otimes \C^d$ is
separable if and only if it is in the closed unit ball of
$\bigotimes_{j=1,\pi}^3 S_1^d$.
\end{thm}

The following lemma will be crucial.
\begin{lem}\label{lem:tensor}
The identity
$$\id:\left(\otimes_{j=1, \Delta_2}^3 \ell_2^d\right)
\otimes_\pi\left(\otimes_{j=1, \Delta_2}^3 \ell_2^d\right) \lra
\otimes_{j=1,\pi}^6\ell_2^d$$ has norm $\succeq d^2$, where
$\Delta_2$ is the usual (Hilbert-Schmidt) tensor norm that makes
$\ell_2^d\otimes_{\Delta_2}\ell_2^d=\ell_2^{d^2}$.
\end{lem}

We will need some concepts about unconditionality on Banach
spaces. We refer the reader to \cite[Chap. 17]{DiJaTo}. First we
need to introduce a couple definitions. Given a linear operator
between two finite dimensional Banach spaces $u:X\lra Y$, we
define its $1$-summing norm $\pi_1(u)$ as the smallest constant
$K$ that makes the following inequality hold for arbitrarily
chosen elements $x_j\in X$:
$$\sum_i\|u(x_i)\|\le K \sup_{x^*\in X^*,\; \|x^*\|\le 1}\sum_i |x^*(x_i)|.$$
We can also define the $1$-factorable norm of $u$, namely
$\gamma_1(u)$, as $ \inf \|a\|\|b\|$, where $a:X\lra \ell_1^N$,
$b:\ell_1^N\lra Y$ and $u=ba$.

Both norms define operator ideals \cite{Defant} in the sense that
they verify the inequalities $\pi_1(uvw)\le \|u\|\pi_1(v)\|w\|$,
$\gamma_1(uvw)\le \|u\|\gamma_1(v)\|w\|$.

\

Now, given a finite dimensional Banach space $X$, we can define its
Gordon-Lewis constant $gl(X)$ as the smallest constant $K$ such that
$\gamma_1(u)\leq k \pi_1(u)$ for every linear operator $u:X\lra
\ell_2^N$.

\begin{proof}{(of Lemma \ref{lem:tensor})}
In \cite{Defant-JFA} it is proven that the Gordon-Lewis constant
$\gl(\otimes_{j=1,\epsilon}^3\ell_2^d)\simeq d$, which by duality
\cite[Prop. 17.9]{DiJaTo} implies
$\gl(\otimes_{j=1,\pi}^3\ell_2^d)\simeq d$.

By the ideal property of $\gamma_1$ and $\pi_1$, it can be easily
deduced that if $u:X\lra Y$ and $v:Y\lra X$ are two operators such
that $id_X=vu$, then $gl(X)\leq \|u\|\|v\|gl(Y)$. Then, since
$\gl(\otimes_{j=1,\Delta_2}^3\ell_2^d)\simeq 1$ \cite[Cor.
4.12]{DiJaTo}, the norm of the identity
$\id:\otimes_{j=1,\Delta_2}^3\ell_2^d \lra
\otimes_{j=1,\pi}^3\ell_2^d$ has to be $\succeq d$. The Lemma is
then a consequence of the metric mapping property for the $\pi$
tensor norm (see Section \ref{sec:preliminaries}).
\end{proof}

Since the trace class $S_1^d$ can be identified with
$\ell_2^d\otimes_\pi\ell_2^d$, Lemma \ref{lem:tensor} implies that
there exists a $d^3\times d^3$ matrix $\rho$ such that
$\|\rho\|_{S_1^{d^3}}=1$ and $\|\rho\|_\pi\simeq d^2$. Using the
Cartesian decomposition $\rho=\re \rho+i \im \rho$ one can assume
$\rho$ to be hermitian, and then, decomposing again into the
positive and negative part, one obtains that $\rho$ can indeed be
taken positive, and hence with trace $1$. But now, the state
$\rho'=p\rho+(1-p)\frac{\1}{d^3}$ verifies that $\|\rho'\|_\pi>1$
and is therefore entangled for every
$p>\frac{2}{1+\|\rho\|_\pi}\simeq\frac{1}{d^2}.$ This proves Theorem
\ref{thm:separable}.

\end{document}